\documentclass{aa}

\usepackage[varg]{txfonts}
\usepackage{natbib}
\usepackage{graphicx}
\usepackage{caption} 
\usepackage{subcaption}
\usepackage{mwe}
\usepackage{longtable}
\usepackage{scalerel}
\usepackage{mathtools}
\usepackage{booktabs} 
\usepackage{multicol} 
\usepackage{multirow} 
\usepackage{array} 
\usepackage{mathabx} 
\usepackage{arydshln}
\usepackage{verbatim} 
\usepackage{lscape} 
\usepackage{pdflscape} 
\usepackage[export]{adjustbox} 
\usepackage{color,epstopdf} 
	\DeclareGraphicsExtensions{.ps}
\usepackage{amsmath}
\newcommand{\RN}[1]{\textup{\uppercase\expandafter{\romannumeral#1}}}
\newcolumntype{L}[1]{>{\raggedright\let\newline\\\arraybackslash\hspace{0pt}}m{#1}}

\begin{document}

\title{Methylamine and other simple N-bearing species in the hot cores NGC 6334\RN{1} MM1 -- 3}

\titlerunning{Methylamine in the hot cores NGC 6334\RN{1} MM1 -- 3}


\author{Eva G. B\o gelund\inst{\ref{inst1}}
	\and Brett A. McGuire\inst{\ref{inst2}}
	\and Michiel R. Hogerheijde\inst{\ref{inst1}, \ref{inst3}}
	\and Ewine F. van Dishoeck\inst{\ref{inst1}, \ref{inst4}}
	\and Niels F. W. Ligterink\inst{\ref{inst1}, \ref{inst5}}
} 


\institute{Leiden Observatory, Leiden University, PO Box 9513, 2300
	RA Leiden, The Netherlands\label{inst1} \newline \email{bogelund@strw.leidenuniv.nl}
	\and National Radio Astronomy Observatory, 520 Edgemont Rd, Charlottesville, VA 22903, USA\label{inst2}
	\and Anton Pannekoek Institute for Astronomy, University of Amsterdam, Science Park 904, 1098 XH Amsterdam, The Netherlands\label{inst3}
	\and Max-Planck Institut für Extraterrestrische Physik, Giessenbachstr. 1, 85748 Garching, Germany\label{inst4}
	\and Center for Space and Habitability (CSH), University of Bern, Sidlerstrasse 5, 3012 Bern, Switzerland\label{inst5}
}

\date{Submitted 19/06/2018, Accepted 17/02/2019}

\abstract
{In the search for the building blocks of life, nitrogen-bearing molecules are of particular interest since nitrogen-containing bonds are essential for the linking of amino acids and ultimately the formation of larger biological structures. The elusive molecule methylamine (CH$_3$NH$_2$) is thought to be a key pre-biotic species but has so far only been securely detected in the giant molecular cloud Sagittarius B2.}
{We identify CH$_3$NH$_2$ and other simple nitrogen-bearing species involved in the synthesis of biologically relevant molecules towards three hot cores associated with the high-mass star-forming region NGC 6334\RN{1}, located at a distance of 1.3 kpc. Column density ratios are derived in order to investigate the relevance of the individual species as precursors of biotic molecules.}
{High sensitivity, high angular and spectral resolution observations obtained with the Atacama Large Millimeter/submillimeter Array were used to study transitions of CH$_3$NH$_2$, CH$_2$NH, NH$_2$CHO, and the $^{13}$C- and $^{15}$N-methyl cyanide (CH$_3$CN) isotopologues, detected towards NGC 6334\RN{1}. Column densities are derived for each species assuming local thermodynamic equilibrium and excitation temperatures in the range 220--340 K for CH$_3$NH$_2$, 70--110 K for the CH$_3$CN isotopologues and 120--215 K for NH$_2$CHO and CH$_2$NH.}
{We report the first detections of CH$_3$NH$_2$ towards NGC 6334\RN{1} with column density ratios with respect to CH$_3$OH of 5.9$\times$10$^{-3}$, 1.5$\times$10$^{-3}$ and 5.4$\times$10$^{-4}$ for the three hot cores MM1, MM2, and MM3, respectively. These values are slightly lower than the values derived for Sagittarius B2 but higher by more than order of magnitude as compared with the values derived for the low-mass protostar IRAS 16293--2422B. The column density ratios of NH$_2$CHO, $^{13}$CH$_3$CN, and CH$_3$C$^{15}$N with respect to CH$_3$OH are (1.5--1.9)$\times$10$^{-4}$, (1.0--4.6)$\times$10$^{-3}$ and (1.7--3.0)$\times$10$^{-3}$ respectively. Lower limits of 5.2, 1.2, and 3.0 are reported for the CH$_3$NH$_2$ to CH$_2$NH column density ratio for MM1, MM2, and MM3 respectively. These limits are largely consistent with the values derived for Sagittarius B2 and higher than those for IRAS 16293--2422B.}
{The detections of CH$_3$NH$_2$ in the hot cores of NGC 6334\RN{1} hint that CH$_3$NH$_2$ is generally common in the interstellar medium, albeit that high-sensitivity observations are essential for the detection of the species. The good agreement between model predictions of CH$_3$NH$_2$ ratios and the observations towards NGC 6334\RN{1} indicate a main formation pathway via radical recombination on grain surfaces. This process may be stimulated further by high grain temperatures allowing a lager degree of radical mobility. Further observations with ALMA will help evaluate the degree to which CH$_3$NH$_2$ chemistry depends on the temperature of the grains in high- and low-mass regions respectively.}

\keywords{Astrochemistry - Methods: observational - Stars: protostars - ISM: individual objects: NGC 6334\RN{1} - Submillimeter: ISM}
 
\maketitle
\section{Introduction} \label{sec:intro}

A number of molecular species that are recognised as precursors to biologically relevant molecules have in recent years been identified in the interstellar medium (ISM). These so-called pre-biotic species \citep[see][and references therein]{Herbst2009} are involved in the formation of, for example, amino acids, the main constituents of proteins, and nucleobases, the fundamental components of DNA and RNA, and thereby constitute the basis for the building blocks of life.

Among the pre-biotic molecules are the species methylamine (CH$_3$NH$_2$) and methanimine (CH$_2$NH), the simplest primary amine- (-NH$_2$) and imine- (-C=N-) containing species, respectively. Experiments in which interstellar ice analogues are subjected to thermal processing or irradiation by UV photons have shown that both CH$_3$NH$_2$ and CH$_2$NH are involved in reactions that form amino acids, and have specifically been proven to engage in the synthesis of glycine (NH$_2$CH$_2$COOH), the smallest member of the amino acid family \citep{Holtom2005, Lee2009, Bossa2009, Danger2011}. The formation of glycine within or upon the icy mantles of interstellar dust-grains is consistent with theoretical models by \cite{Garrod2013} who trace and couple the gas-phase, grain-surface and bulk ice chemistry during the formation of hot cores. In addition, the connection between CH$_3$NH$_2$ and glycine has been established though the proposed formation of both these species from a common set of precursors present in carbonaceous chondrite meteorites \citep{Aponte2017} including carbon monoxide (CO), ammonia (NH$_3$), hydrogen cyanide (HCN), and carbon dioxide (CO$_2$).

Another example of a simple progenitor of biotic molecules is formamide (NH$_2$CHO), the simplest amide (-NH-(C=O)-), which has the same chemical structure as the peptide bonds that link amino acids and thereby form the backbone of larger protein structures. NH$_2$CHO has also been shown to be involved in the formation of nucleobases and nucleobase analogues in processes which use minerals and metal oxides, including samples of primitive meteoroids, as catalysts \citep{Saladino2006, Kumar2014, Saladino2016}. 

Lastly, due to its cyanide (-CN) group, the molecule methyl cyanide (acetonitrile, CH$_3$CN) is also of interest in relation to the synthesis of pre-biotic molecules. This is due to the importance of C-N bonds for the formation of peptide structures. Reactions involving cyanides, especially HCN and its derivatives, are therefore regarded as the foundation of the formation of complex structures such as proteins, lipids and nuclei acids \citep{Matthews2006, Patel2015}. In addition, \cite{Goldman2010} propose that shock-induced C-N bonds due to cometary impacts on the early Earth provide a potential synthesis route for amino acids which is independent of the pre-existing atmospheric conditions and materials on the planet. In summary, continued observations and searches for CH$_3$NH$_2$, CH$_2$NH, NH$_2$CHO, CH$_3$CN, and other pre-biotic species in the ISM, as well as in solar system bodies, are of high interest in order to establish the relevance of the respective species in connection to the emergence of life on Earth, and potentially on other (exo)planets and moons.


NH$_2$CHO and CH$_3$CN are routinely detected towards high- and low-mass hot cores \citep{Cazaux2003, Bisschop2007, Kahane2013}, and have in addition been identified towards a number of comets \citep[see review by][]{Mumma2011}, in particular the bright comet Hale-Bopp \citep[e.g.][]{Bockelee-Morvan1997, Remijan2008} and comet 67P/Churyumov–Gerasimenko (hereafter 67P), the target of ESA's \textit{Rosetta} mission \citep{Goesmann2015, Altweeg2017}. In addition, CH$_3$CN was the first complex organic molecule (COM) to also be detected in a protoplanetary disk \citep{Oberg2015} and thereby became one of the few pre-biotic species whose presence could be traced throughout all formation phases from the earliest stages of star-formation to the last remnants in comets. 

Despite the lack of firm detections of CH$_2$NH in comets \citep{Irvine1998, Crovisier2004}, this species has also been detected towards a variety of interstellar sources including giant molecular clouds \citep{Dickens1997} and high- and low-mass protostellar systems \citep{Suzuki2016, Ligterink2018}. In contrast to these detections, the structurally similar species CH$_3$NH$_2$ has proven to be an especially elusive molecule and for a long time was only securely detected towards the high-mass source Sagittarius B2 (hereafter Sgr B2) located in the Galactic centre \citep[e.g.][]{Kaifu1974, Belloche2013}. Recently, the molecule was also detected towards the hot core G10.47+0.03 by \citet{Ohishi2017} who also report a tentative detection towards NGC 6334\RN{1} though the low signal-to-noise and variations in $v_{\textrm{LSR}}$ between transitions of the species makes the detection unclear. A tentative detection was also reported towards Orion KL by \citet{Pagani2017}. In addition, a series of non-detections have been reported towards a number of high-mass young stellar objects \citep[YSOs,][]{Ligterink2015} and a very stringent upper limit has been set on the abundance of the species in the low-mass Sun-like protostar IRAS 16293--2422B \citep{Ligterink2018}. Recently, the species has also been detected in the coma of comet 67P \citep{Altweeg2017}. These detections (and upper limits) indicate a range of CH$_3$NH$_2$ abundances with respect to CH$_3$OH, with that of IRAS 16293--2422B being at least one to two orders of magnitude lower than the values derived for Sgr~B2. The discrepancies between the detections in Sgr B2 and the non-detections elsewhere has led to the suggestion that formation pathways for CH$_3$NH$_2$ are not very efficient and that they may depend strongly on the conditions which characterise the individual regions. Based on the detections of CH$_3$NH$_2$ in Sgr B2 it has therefore been speculated that the presence of relatively high dust grain temperatures or strong radiation fields enhance CH$_3$NH$_2$ formation. 

The formation of CH$_3$NH$_2$ is discussed in a number of studies. On interstellar dust grains, two main formation pathways have been proposed: The first is a hydrogenation sequence starting from hydrogen cyanide: HCN + 2H $\rightarrow$ CH$_2$NH + 2H $\rightarrow$ CH$_3$NH$_2$ \citep{Theule2011}. Although the efficiency of formation via this pathway is ill constrained, the same hydrogenation mechanism has been used in glycine formation models to form the intermediate CH$_2$NH$_2$ radical \citep{Woon2002}. The second formation route involves radical recombination reactions between a methyl (-CH$_3$) and an amino group: CH$_3$ + NH$_2$ $\rightarrow$ CH$_3$NH$_2$. This pathway has been included in the astrochemical models presented by \cite{Garrod2008} as the main formation route for CH$_3$NH$_2$. Experimentally, electron and photon irradiated interstellar ice analogues, consisting of CH$_4$ and NH$_3$, have been shown to result in formation of CH$_3$NH$_2$ \citep{Kim2011, Forstel2017}. Though in dark clouds, both CH$_3$ and NH$_2$ can also result from H-addition to atomic C and N and therefore photodissociation is not critical for the formation of the radicals. In the gas-phase, the radical-neutral reaction CH$_3$ + NH$_3$ $\rightarrow$ CH$_3$NH$_2$ + H has been proposed to be the main CH$_3$NH$_2$ formation route. This is based on the observational study of Sgr B2 conducted by \cite{Halfen2013} who also argue that the formation of CH$_3$NH$_2$ through successive hydrogenation of CH$_2$NH is unlikely due to the large difference in rotational temperature, 44$\pm$13 K in the case of CH$_2$NH and 159$\pm$30 K in the case of CH$_3$NH$_2$, derived through rotational temperature diagrams. This difference makes it unlikely that the molecules occupy the same regions thereby making CH$_2$NH an unlikely synthetic precursor of CH$_3$NH$_2$. A dominant gas-phase formation route for CH$_2$NH is also reported by \citet{Suzuki2016} though they note that hydrogenation of solid-phase CH$_2$NH can also form CH$_3$NH$_2$. Additional detections of CH$_3$NH$_2$ and related species, preferably towards a large number of different sources, will therefore provide valuable information and help distinguish between formation routes and conditions required for the formation of this species.

In this work, CH$_3$NH$_2$ along with other simple pre-biotic nitrogen-bearing species, in particular CH$_2$NH, CH$_3$CN and NH$_2$CHO, are studied towards three dense cores within the giant molecular cloud complex NGC 6334. The NGC 6334 region, located in the constellation Scorpius in the southern hemisphere, is a very active high-mass star-forming region composed of six sub-regions denoted \RN{1}--\RN{5} and \RN{1}(N) \citep[see review by][and references therein]{Persi2008}. Water and methanol (CH$_3$OH) maser studies have placed the region at a mean distance of 1.3 kpc from the Sun \citep{Chibueze2014, Reid2014}, equivalent to a galactocentric distance ($d{_\textrm{GC}}$) of $\sim$7.02 kpc. The focus of this work is on the deeply embedded source NGC 6334\RN{1} which is located in the north-eastern part of the cloud. The morphology of this source has been studied in detail by \cite{Brogan2016} who identify a number of distinct peaks in the sub-millimetre continuum and assign these to individual high-mass star-forming systems. The region has a very rich molecular inventory as demonstrated by \cite{Zernickel2012} who identify a total of 46 molecular species towards NGC6334\RN{1} including CH$_2$NH, CH$_3$CN, and NH$_2$CHO but not CH$_3$NH$_2$. 

This paper presents the first detection of CH$_3$NH$_2$ towards NGC 6334\RN{1}. The work is based on high sensitivity, high spectral and angular resolution data obtained with the Atacama Large Millimeter/submillimeter Array (ALMA). Previous searches for CH$_3$NH$_2$ have, for the most part, been carried out with single dish telescopes, which are generally less sensitive when compared with interferometric observations, and have therefore focused mainly on the bright hot cores associated with the Galactic central region. With the unique sensitivity and resolving power of ALMA this is changing and the weak lines associated with CH$_3$NH$_2$ can now be probed in regions away from the Galactic centre, such as NGC 6334\RN{1}, as well as in low-mass systems \citep{Ligterink2018}. 

The paper is structured in the following way: in Sect. \ref{sec:Method} the observations and analysis methodology are introduced. Section \ref{sec:results} presents the observed transitions of each of the studied species and the model parameters used to reproduce the data. In Sect. \ref{sec:discuss} the derived column density ratios are discussed and compared between the regions in NGC 6334\RN{1} as well as to the values derived for other high- and low-mass objects. Finally, our findings are summarised in Sect. \ref{sec:conclusion}. 

\section{Observations and method} \label{sec:Method}
\subsection{Observations} \label{subsec:obs}
Observations of NGC 6334\RN{1} were carried out with ALMA in Cycle 3 on January 17, 2016 using the ALMA Band 7 receivers (covering the frequency range 275--373 GHz). Three spectral windows centred around 301.2, 302.0, and 303.7 GHz covering a total bandwidth of $\sim$3 GHz were obtained. The observations have spectral and angular resolutions of 1 km s$^{-1}$ and $\sim$1$\arcsec$ (equivalent to $\sim$1300 au at the distance of NGC 6334\RN{1}) respectively. The data were interactively self-calibrated and continuum subtracted using the most line-free channels. A detailed description of this reduction procedure may be found in \cite{Brogan2016} and \cite{Hunter2017} while a summary of all observing parameters are listed in Table 1 of \cite{McGuire2017}. After calibration the data were corrected for primary beam attenuation.

\subsection{Method} \label{subsec:method}
\begin{figure}
	\centering
	\includegraphics[width=0.48\textwidth]{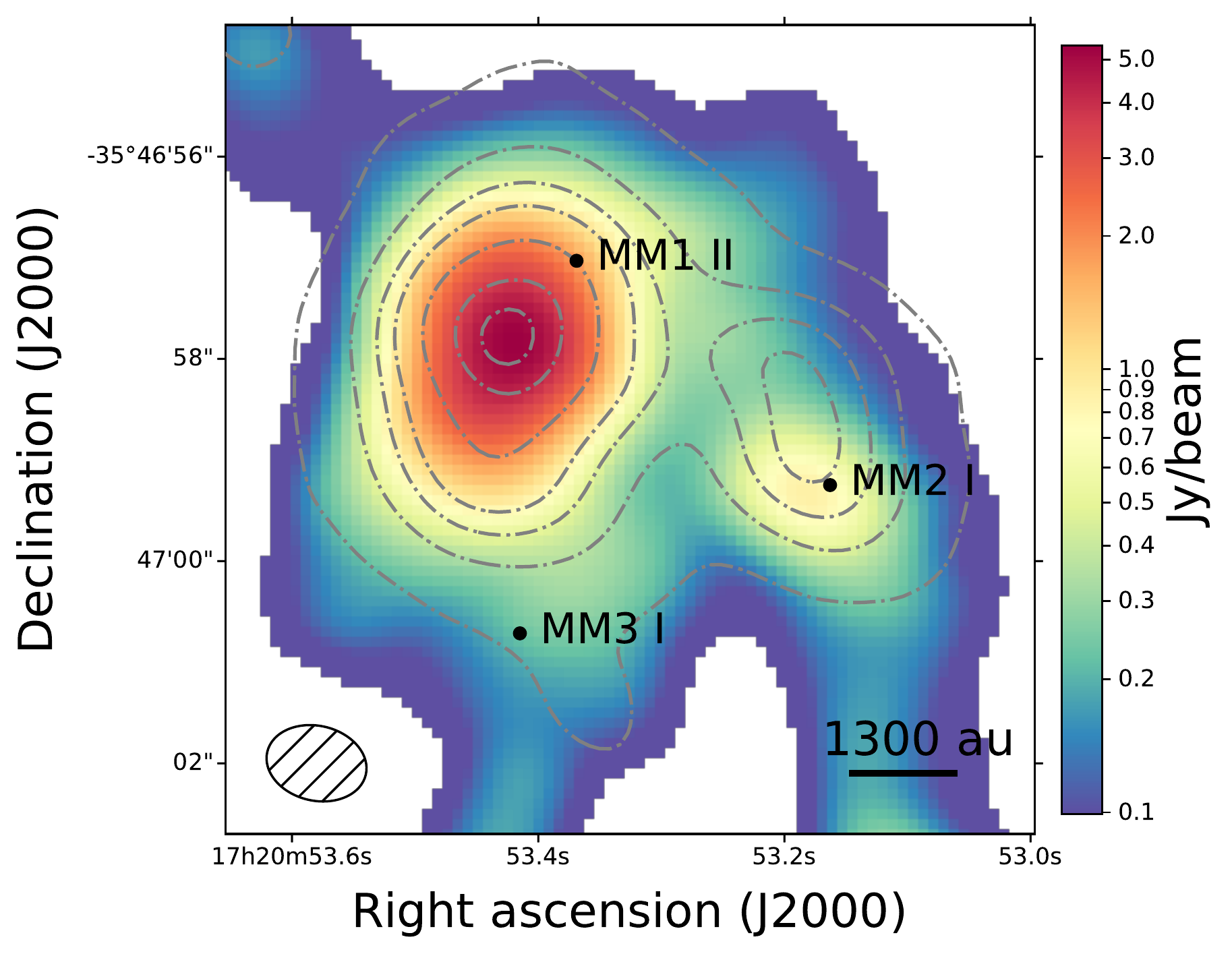}
	\caption[]{1 mm continuum image of the NGC 6334\RN{1} region with the velocity integrated intensity map of the $^{13}$CH$_3$CN transition at 303.610 GHz overlaid in grey contours (levels are [3, 20, 40, 60, 100, 150, 180]$\sigma$ with $\sigma$=0.07 Jy beam$^{-1}$ km s$^{-1}$). Pixels with values less than 1\% of the peak intensity have been masked out. The locations at which spectra have been extracted are marked for each region. The synthesised beam ($\sim$1300$\times$962~au) is shown in the bottom left corner.}
	\label{fig:Map}
\end{figure}

For the analysis of CH$_3$NH$_2$ and related species three spectra, extracted at different locations across the NGC6334\RN{1} region, are used. For consistency we use the same locations and naming as in \cite{Bogelund2018} and focus on the regions MM1~\RN{2}, MM2~\RN{1}, and MM3~\RN{1}. These regions are associated with each of the continuum sources MM1, MM2, and MM3 making it possible to compare the abundances of the various species across the three hot cores. Due to the greater lines widths characterising the central part of the MM1 region and the bright continuum emission, which in some cases result in negative features after continuum subtraction has been applied, we select a region away from the main continuum peak where weak emission line features are more easily identified. The extracted spectra are the average of a 1$\overset{\second}{.}$00$\times$0$\overset{\second}{.}$74 region, equivalent to the area of the synthesised beam. The coordinates of the central pixel of each of the regions are (J2000 17$^{\rm{h}}$20$^{\rm{m}}$53.371$^{\rm{s}}$, $-35^{\circ}$46\arcmin57.013\arcsec), (J2000~17$^{\rm{h}}$20$^{\rm{m}}$53.165$^{\rm{s}}$, $-35^{\circ}$46\arcmin59.231\arcsec) and (J2000 17$^{\rm{h}}$20$^{\rm{m}}$53.417$^{\rm{s}}$, $-35^{\circ}$47\arcmin00.697\arcsec) for MM1~\RN{2}, MM2~\RN{1}, and MM3~\RN{1} respectively. For each of the extracted spectra, the rms noise is calculated after careful identification of line-free channels. These are $\sim$0.9 K (68 mJy beam$^{-1}$) for MM1, $\sim$0.6 K (45 mJy beam$^{-1}$) for MM2, and $\sim$0.04 K (3 mJy beam$^{-1}$) for MM3. The difference in the estimated rms noise values reflects the large variations in brightness and line density over the three regions. An overview of the NGC 6334 \RN{1} region and the locations at which spectra have been extracted is shown in Fig. \ref{fig:Map}.

In order to identify transitions of CH$_3$NH$_2$, CH$_2$NH, CH$_3$CN, and NH$_2$CHO, as well as to constrain the column density and excitation temperature of the species at each of the studied positions, synthetic spectra are produced using the CASSIS\footnote{Centre d'Analyse Scientifique de Spectres Instrumentaux et Synthétiques; http://cassis.irap.omp.eu} line analysis software. The spectroscopic data for CH$_2$NH and the methyl cyanide and formamide isotopologues are adopted from the JPL\footnote{Jet Propulsion Laboratory \citep{Pickett1998}; http://spec.jpl.nasa.gov} and CDMS\footnote{Cologne Database for Molecular Spectroscopy \citep{Muller2001,Muller2005}; http://www.ph1.uni-koeln.de/cdms/} molecular databases. For CH$_3$NH$_2$, the spectroscopic data are taken from \citet{Motiyenko2014}. Assuming local thermodynamic equilibrium (LTE) and optically thin lines, synthetic spectra are constructed for each species. This is done by providing CASSIS with a list of parameters including excitation temperature, $T_{\textrm{ex}}$ [K], column density of the species, $N_{\textrm{s}}$ [cm$^{-2}$], source velocity, $v_{\textrm{LSR}}$ [km s$^{-1}$], line width at FWHM [km s$^{-1}$], and angular size of the emitting region, $\theta_{\textrm{s}}$ [$^{\second}$], assumed to be equal to the size of the synthesised beam. 

Excitation temperatures and column densities are determined for the detected species by creating grids of model spectra varying $T_{\textrm{ex}}$ and $N_{\textrm{s}}$ and identifying the model spectrum with the minimal $\chi^2$ as the best fit. The CASSIS software computes the $\chi^2$ value for each of the model spectra taking into account the rms noise of the observed spectrum and the calibration uncertainty (assumed to be $\sim$10\%). $\chi^2$ is defied as:

\begin{equation}
\chi^2 = \sum_{i = 1}^{N} \left(\frac{(I_{\textrm{obs},i} - I_{\textrm{model},i})}{\sigma_i}\right)^2,
\end{equation}

\noindent where $I_{\textrm{obs},i}$ and $I_{\textrm{model},i}$ is the intensity of the observed and modelled spectrum in channel $i$, respectively, \linebreak $\sigma_i = \sqrt{rms^2 + (0.1 \times I_{\textrm{obs},i})^2}$ and $N$ is the number of fitted points, that is, the number of channels covered by each of the transitions to which the model is optimised (we consider the channels within a range of $\pm~2~\times~FWHM$). Table~\ref{tab:model_grid} lists the model grids for each of the fitted species. Since only a single CH$_2$NH transition and just two NH$_2$CHO transitions are detected, the excitation temperatures of these species cannot be constrained from the data. The reported CH$_2$NH and NH$_2$CHO column densities are therefore derived assuming $T_{\textrm{ex}}$ to be fixed at the value derived for CH$_3$OH in each region by \cite{Bogelund2018}. These are 215 K for region MM1~\RN{2}, 165 K for region MM2~\RN{1}, and 120 K for region MM3~\RN{1}. The uncertainty on $N_{\textrm{s}}$ and $T_{\textrm{ex}}$ is listed as the standard deviation of model spectra with $\chi^2$ within 1$\sigma$ of the best-fit model. For $N_{\textrm{s}}$, the highest uncertainty is approximately 30\% while the uncertainty on $T_{\textrm{ex}}$ is up to 65\%. Through the propagation of errors, the uncertainty on listed column density ratios is conservatively estimated to be $\sim$40\% ($\sqrt{2}\times30\%$). Because the velocity structure of NGC 6334\RN{1} is not well-known, the  source velocity and FWHM line widths characterising each region are fixed throughout the fitting procedure so as not to introduce additional free parameters. As is clear from Fig.~\ref{fig:CH3NH2lines} and Figs.~\ref{fig:13CH3CNlines}~--~\ref{fig:NH2CHOlines}, the fixed $v_{\textrm{LSR}}$ and line widths are consistent with the data for all species. However, examples of molecules detected towards the same region but characterised by different physical parameters have been reported \citep[see e.g.][]{Halfen2013}.

For each identified CH$_3$NH$_2$, CH$_2$NH, CH$_3$CN, or NH$_2$CHO transition, a thorough search for potential blending species is conducted. This search is carried out carefully in the following steps: 1) All catalogued species, that is to say all species which are listed in the JPL or CDMS databases and which have transitions at frequencies that overlap with those of CH$_3$NH$_2$, CH$_2$NH, CH$_3$CN, or NH$_2$CHO, are identified. 2) For each potential blending species a synthetic spectrum is produced and optimised so that the column density of that species is maximised. This is done while ensuring that none of the other transitions belonging to the same species, and which are covered by the data, are overproduced with respect to the data. 3) If the potential blending species are isotopologues, step 2 is repeated for the parent species in order to ensure that column densities are consistent between isotopologues of the same species. 4) Once the spectra of the individual potential blending species have been optimised, they are summed to obtain a full spectrum for each of the three regions. Two fits are then preformed; the first fit takes only the studied species into account and is used to set an upper limit on the column density for each of these; the second fit includes the contributions from all potential blending species.

By including the maximised contribution from the potential blending species to the modelled spectrum, the contributions from CH$_3$NH$_2$, CH$_2$NH, CH$_3$CN, and NH$_2$CHO to the same modelled spectrum are minimised and consequently the most stringent limits on the column densities of these species are achieved. It should be noted however, that maximising the column densities of some potential blending species, in particular deuterated isotopologues, leads to values which are unrealistically high when compared with parent species and therefore should be seen purely as a method to conservatively constrain the amounts of CH$_3$NH$_2$, CH$_2$NH, CH$_3$CN, and NH$_2$CHO. The full list of potential blending species as well as model parameters are listed in Table \ref{tab:main_blending_species}. Finally, a $^{12}$C/$^{13}$C ratio of 62, a $^{16}$O/$^{18}$O ratio of 450, and a $^{14}$N/$^{15}$N ratio of 422 is adopted throughout the paper, all derived assuming $d_{\textrm{GC}}$ = 7.02 kpc and the relations for $^{12}$C/$^{13}$C, $^{16}$O/$^{18}$O and $^{14}$N/$^{15}$N reported by \cite{Milam2005} and \cite{Wilson1999}.

\section{Results} \label{sec:results}
In the following sections the detections of CH$_3$NH$_2$ will be discussed in detail alongside a summary of the main results regarding the detections of CH$_2$NH, CH$_3$CN, and NH$_2$CHO (see Appendix \ref{app:results} for full discussion of these species). Transition frequencies and line data for all species are listed in Table \ref{tab:line_summary}, while integrated line intensities of a select number of lines in the observational data are listed in Table \ref{tab:integrated_lines}. In the case of CH$_3$NH$_2$, detected lines have $E_{\textrm{up}}$ values ranging from 96 to 480 K. For each of the studied regions and species the column density and excitation temperature of the best-fit synthetic spectrum are derived. In Sect. \ref{sec:discuss} these values and their ratios with respect to CH$_3$OH and CH$_3$NH$_2$ will be compared between the individual regions of NGC 6334\RN{1} but also discussed in relation to those derived for other objects. The parameters of the best-fit models are listed in Table~\ref{tab:models} and all transitions and modelled spectra of CH$_3$NH$_2$ and other species are plotted in Fig.~\ref{fig:CH3NH2lines} and Figs.~\ref{fig:13CH3CNlines}~--~\ref{fig:NH2CHOlines} respectively.

\begin{table*}[]
	\centering
	\caption{Summary of lines}
	\label{tab:line_summary}
	\begin{tabular}{cccccc}
		\toprule
		\multicolumn{2}{c}{Transition} & Catalogue Frequency & $E_{\textrm{up}}$ & $A_{\textrm{ij}}$ & Catalogue \\
		$[\textrm{QN}]_{\textrm{up}}$\tablefootmark{a} & $[\textrm{QN}]_{\textrm{low}}$\tablefootmark{a} & [MHz] & [K] & $\times$10$^{-5}$ [s$^{-1}$] & \\
		\midrule
		\multicolumn{6}{c}{$\rm{CH_2NH}$} \\
		15$_{2,13}$ 14 & 14$_{3,12}$ 13 & 302 565.4318 & 408.72 & 6.61 & JPL \\
		15$_{2,13}$ 16 & 14$_{3,12}$ 15 & 302 565.4883 & 408.72 & 6.64 & \\
		15$_{2,13}$ 15 & 14$_{3,12}$ 14 & 302 566.3219 & 408.72 & 6.61 & \\
		\multicolumn{6}{c}{$\rm{CH_3NH_2}$\tablefootmark{b}} \\
		16 2 A2 15 & 15 3 A1 14 & 301 247.6939 & 305.21 & 1.37 & \citet{Motiyenko2014} \\
		16 2 A2 17 & 15 3 A1 16 & 301 247.7074 & 305.21 & 1.55 & \\
		16 2 A2 16 & 15 3 A1 16 & 301 247.9700 & 305.21 & 1.46 & \\		
		13 2 B2 13 & 13 1 B1 13 & 301 424.0139 & 210.13 & 2.69 & \\
		13 2 B2 14 & 13 1 B1 14 & 301 425.6883 & 210.13 & 2.90 & \\
		13 2 B2 13 & 13 1 B1 12 & 301 425.8175 & 210.13 & 2.50 & \\
		9 0 B2 8 & 8 1 B1 7 & 301 653.3284 & 95.93 & 2.68 & \\
		9 0 B2 10 & 8 1 B1 9 & 301 653.4789 & 95.93 & 3.36 & \\
		9 0 B2 9 & 8 1 B1 8 & 301 654.7988 & 95.93 & 3.00 & \\
		16 7 B1 16 & 17 6 B2 17 & 302 801.6275 & 480.13 & 0.83 & \\
		16 7 B2 16 & 17 6 B1 17 & 302 801.6306 & 480.15 & 0.83 & \\
		16 7 B1 17 & 17 6 B2 18 & 302 801.7834 & 480.13 & 0.88 & \\
		16 7 B2 17 & 17 6 B1 18 & 302 801.7866 & 480.13 & 0.88 & \\
		16 7 B1 15 & 17 6 B2 16 & 302 801.7912 & 480.13 & 0.78 & \\
		16 7 B2 15 & 17 6 B1 16 & 302 801.7943 & 480.13 & 0.78 & \\
		9 0 E2+1 8 & 8 1 E2+1 7 & 303 733.9183 & 96.20 & 2.63 & \\
		9 0 E2+1 10 & 8 1 E2+1 9 & 303 734.0611 & 96.20 & 3.29 & \\
		9 0 E2+1 9 & 8 1 E2+1 8 & 303 735.3214 & 96.20 & 2.94 & \\
		\multicolumn{6}{c}{$\rm{^{13}CH_3CN}$} \\
		17$_{5}$ & 16$_{5}$ & 303 518.8535 & 310.00 & 222 & CDMS \\
		17$_{4}$ & 16$_{4}$ & 303 570.0991 & 245.64 & 230 & \\
		17$_{3}$ & 16$_{3}$ & 303 609.9710 & 195.57 & 236 & \\
		17$_{2}$ & 16$_{2}$ & 303 638.4820 & 159.80 & 240 & \\
		17$_{1}$ & 16$_{1}$ & 303 655.5770 & 138.33 & 243 & \\
		17$_{0}$ & 16$_{0}$ & 303 661.2780 & 131.18 & 243 & \\
		\multicolumn{6}{c}{$\rm{CH_3C^{15}N}$} \\
		17$_{4}$ & 16$_{4}$ & 303 187.8887 & 245.49 & 229 & \\
		17$_{3}$ & 16$_{3}$ & 303 227.9360 & 195.41 & 235 & \\
		17$_{2}$ & 16$_{2}$ & 303 256.5540 & 159.64 & 240 & \\
		17$_{1}$ & 16$_{1}$ & 303 273.7300 & 138.17 & 242 & \\
		17$_{0}$ & 16$_{0}$ & 303 279.4560 & 131.01 & 243 & \\
		\multicolumn{6}{c}{$\rm{NH_2CHO}$} \\
		15$_{1,15}$ & 14$_{1,14}$ & 303 450.2040 & 120.01 & 205 & CDMS \\
		14$_{1,13}$ & 13$_{1,12}$ & 303 660.5390 & 113.01 & 204 & \\
		\multicolumn{6}{c}{$\rm{NH_2^{13}CHO}$} \\
		15$_{1,15}$ & 14$_{1,14}$ & 302 553.9861 & 119.61 & 203 & CDMS \\
		14$_{1,13}$ & 13$_{1,12}$ & 303 111.8280 & 112.78 & 203 & \\
		\bottomrule
	\end{tabular}
	\tablefoot{\tablefoottext{a}{Quantum numbers for CH$_2$NH are J$_{\textrm{Ka,Kc}}$~F. Quantum numbers for CH$_3$NH$_2$ are J $\textrm{K}_{\textrm{a}}$ $\Gamma$ F, following the notation of \citet{Motiyenko2014}. Quantum numbers for $^{13}$CH$_3$CN and CH$_3$C$^{15}$N are J$_{\textrm{K}}$. Quantum numbers for NH$_2$CHO and NH$_2^{13}$CHO are J$_{\textrm{Ka,Kc}}$.} \tablefoottext{b}{Only lines with $A_{\textrm{ij}}>10^{-6} s^{-1}$ are listed.}}
\end{table*}

\begin{table*}[]
	\centering
	\caption{Best-fit model parameters}
	\label{tab:models}
	\begin{tabular}{!{\extracolsep{4pt}}lcccccc!{}}
		\toprule
		& \multicolumn{2}{c}{MM1~\RN{2}} & \multicolumn{2}{c}{MM2~\RN{1}} & \multicolumn{2}{c}{MM3~\RN{1}} \\
		\midrule
		$v_{\textrm{LSR}}$ [km s$^{-1}$] & \multicolumn{2}{c}{[-6.7]} & \multicolumn{2}{c}{[-9.0]} & \multicolumn{2}{c}{[-9.0]} \\
		FWHM [km s$^{-1}$] & \multicolumn{2}{c}{[3]} & \multicolumn{2}{c}{[3.5]} & \multicolumn{2}{c}{[3]} \\
		$\theta_{\textrm{s}}$ [$^{\second}$] & \multicolumn{2}{c}{[1]} & \multicolumn{2}{c}{[1]} & \multicolumn{2}{c}{[1]} \\
		\midrule
		& $T_{\textrm{ex}}$ & $N_{\textrm{s}}$ & $T_{\textrm{ex}}$ & $N_{\textrm{s}}$ & $T_{\textrm{ex}}$ & $N_{\textrm{s}}$ \\
		& [K] & [cm$^{-2}$] & [K] & [cm$^{-2}$] & [K] & [cm$^{-2}$] \\ 
		\cline{2-3}
		\cline{4-5}
		\cline{6-7}
		CH$_2$NH 
		& [215] & $\leq$5.2$\times10^{16}$ & [165] & $\leq$5.0$\times10^{16}$ & [120] &$\leq10^{15}$ \\
		CH$_3$NH$_2$ & 340 $\pm$ 60 & (2.7 $\pm$ 0.4) $\times10^{17}$ & 230 $\pm$ 30 & (6.2 $\pm$ 0.9) $\times10^{16}$ & 220 $\pm$ 30 & (3.0 $\pm$ 0.6) $\times10^{15}$ \\ 
		$^{13}$CH$_3$CN & 70 $\pm$ 10 & (3.4 $\pm$ 1.0) $\times10^{15}$ & 80 $\pm$ 25 & (1.4 $\pm$ 0.5) $\times10^{15}$ & 90 $\pm$ 15 & (9.0 $\pm$ 0.8) $\times10^{13}$ \\
		CH$_3$C$^{15}$N & 110 $\pm$ 50 & (3.3 $\pm$ 0.5) $\times10^{14}$ & [80] & (1.8 $\pm$ 0.4)$\times10^{14}$ & 70 $\pm$ 45 & (2.3 $\pm$ 0.7) $\times10^{13}$ \\
		NH$_2$CHO 
		& [215] & (7.0 $\pm$ 1.7) $\times10^{15}$ & [165] & (7.6 $\pm$ 0.8) $\times10^{15}$ & [120] & $\leq$5.0$\times10^{13}$ \\
		NH$_2^{13}$CHO\tablefootmark{a} & [215] & $\leq$2.0$\times10^{15}$ & [165] & $\leq$5.0$\times10^{14}$ & -- & -- \\
		\bottomrule
	\end{tabular}
	\tablefoot{Values in square brackets are fixed. Excitation temperatures for CH$_3$NH$_2$ and CH$_3$CN are the values of the best-fit respective models while $T_{\textrm{ex}}$ for CH$_2$NH and NH$_2$CHO is fixed at the best-fit model value derived for CH$_3$OH \citep{Bogelund2018}. In the MM2 region, the excitation temperature for CH$_3$C$^{15}$N is not well constrained and is therefore adopted from $^{13}$CH$_3$CN. Listed uncertainties are the standard deviation of models with $\chi^2$ within 1$\sigma$ of the best-fit model.} 
\end{table*}

\subsection{Methylamine CH$_3$NH$_2$} \label{subsec:CH3NH2}
\begin{figure*}
	\centering
	\includegraphics[width=1.\textwidth, trim={0 0.5cm 0.5cm 0}, clip]{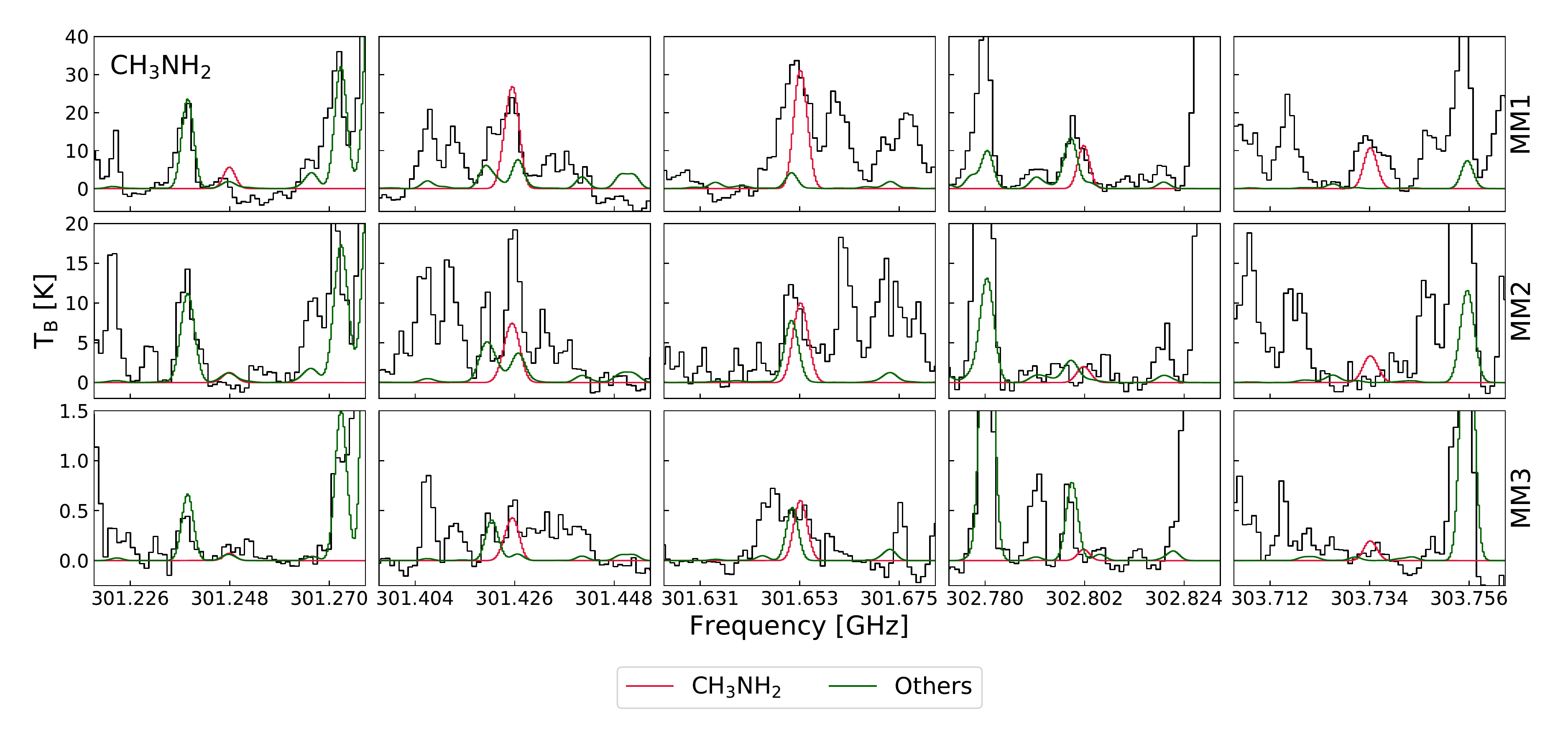} 
	\caption[]{CH$_3$NH$_2$ transitions detected towards NGC6334\RN{1}. Red and green lines represent the synthetic spectrum of CH$_3$NH$_2$ and the sum of spectra of other contributing species respectively. The abscissa is the rest frequency with respect to the radial velocity towards each of the hot cores (listed in Table \ref{tab:models}). 
		The data are shown in black. \textit{Top panels}: MM1~\RN{2}. \textit{Middle panels}: MM2~\RN{1}. \textit{Bottom panels}: MM3~\RN{1}.}
	\label{fig:CH3NH2lines}
\end{figure*}

For CH$_3$NH$_2$, five transitional features (all covering multiple hyperfine components) are identified towards NGC6334\RN{1}. These are plotted in Fig. \ref{fig:CH3NH2lines}. The CH$_3$NH$_2$ transitions are not isolated lines but blended with transitions of other species. Nevertheless, and despite the contributions from the potential blending molecules, it is evident that the data cannot be reproduced without including CH$_3$NH$_2$ in the model, especially for the MM1~\RN{2} and MM3~\RN{1} regions. 

\textbf{MM1~\RN{2}:} For MM1~\RN{2} the CH$_3$NH$_2$ transitions are well reproduced by a model with a column density of 2.7$\times10^{17}$ cm$^{-2}$ and an excitation temperature of 340 K. The uncertainty on each of these values is less than 20\%. For lower excitation temperatures, down to 100 K, the column density is consistent with that derived for 340 K within a factor of approximately two. The same is true for $T_{\textrm{ex}}$ up to 500 K though for very low temperatures, down to 50 K, the column density can no longer be well-constrained. Also, since the variation between the column density of the fit which only takes into account CH$_3$NH$_2$ and the fit which includes all potential blending species is less then 30\%, we consider it very probable that the features in the spectrum of this region are due to CH$_3$NH$_2$. The fact that the features cannot be reproduced without including CH$_3$NH$_2$ in the model makes the detection even more convincing. Around the transition located at 301.248 GHz, a slight negative offset in the baseline is seen. This is likely caused by continuum over-subtraction resulting in a negatively displaced baseline which makes the model transition at this location appear brighter than the observed one.

\textbf{MM2~\RN{1}:} The best-fit model for region MM2~\RN{1} has a column density equal to 6.2$\times10^{16}$ cm$^{-2}$ and an excitation temperature of 230 K. This model is optimised to fit all of the covered CH$_3$NH$_2$ transitions, although only two of these, located at 301.426 GHz and 301.653 GHz, are considered fully detected. The remaining transitions, located at 301.248~GHz, 302.802 GHz, and 303.734 GHz, are considered tentative detections. This is because these transitions are blended with emission from other species (lines at 301.248 and 302.802GHz) or because no clear line is visible in the observed spectrum at the expected location (line at 303.734 GHz). The tentative detections are included in the $\chi^2$ minimisation, as they help constrain the best-fit model. For MM2, the uncertainty on $N_{\textrm{s}}$ and $T_{\textrm{ex}}$ is $\sim$15\%. Varying the excitation temperature down to 50 and up to 500 K does not cause the value of the column density to change by more than a factor of two with respect to the best-fit value derived at 230 K. In contrast to the CH$_3$NH$_2$ features of MM1~\RN{2} however, which are all well reproduced by the single-density, single-temperature model, the lines of MM2~\RN{1} are not. Particularly the line ratio of the transitions at 301.426~GHz and 301.653~GHz is off and cannot be reproduced by the model. Despite the fact that the upper state energy of the transitions is fairly different, $\sim$210~K for the 301.426~GHz transitions and $\sim$96~K for the 301.653 GHz transitions, introducing a two-component model to account for a warm and cool emission region respectively, does not improve the fit. While the transition at 301.653 GHz may be well reproduced by a model with an excitation temperature of $\sim$50 K, addition of any higher excitation temperature-components to the model results in modelled line intensities that vastly overshoot the transition at 301.653 GHz with respect to the data while the intensity of the lines at 301.426 GHz remains much weaker than the observed line. The behaviour of this last transition is especially puzzling since none of the species included in either the JPL or CDMS catalogues are able to reproduce the observed data feature. One possible explanation is of course that the feature in the spectrum of MM2~\RN{1} is due to transitions of some unknown species (or unknown transition of a known species) which is not included in the spectroscopic databases. However, if that is the case, this unknown species is particular to the MM2~\RN{1} region and does not significantly affect regions MM1~\RN{2} and MM3~\RN{1} where the respective CH$_3$NH$_2$ models correspond well with the observations.

The dissimilarity between the CH$_3$NH$_2$ model spectrum and the observations could also indicate that the critical density for individual transitions in the MM2~\RN{1} region may not be reached, removing the region from LTE. Thus, a scenario in which the density of region MM2 is so low that the critical density of one transition is reached, while that of another transition is not, could explain why the model predictions are not able to reproduce the CH$_3$NH$_2$ transitions at 301.425 and 301.655 GHz simultaneously in this region while the same lines are well-matched with the data for regions MM1 and MM3. To test this hypothesis, the collisional coefficients need to be known and the critical densities inferred for each of the transitions in question. However, since these numbers are not known for CH$_3$NH$_2$ we are unable to make the comparison but can instead conclude that it is likely that the MM2 region has a lower overall density as compared with the regions MM1 and MM3. A lower density of the MM2 region with respect to the MM1 region is consistent with the findings of \cite{Brogan2016}, who estimate the dust mass associated with each of the hot cores based on their spectral energy distribution. As in the case of MM1~\RN{2}, the CH$_3$NH$_2$ features cannot be reproduced satisfactory by any other species and therefore we conclude that CH$_3$NH$_2$ is likely to be present in the region despite the inadequacy of the model to fully reproduce the data.

\textbf{MM3~\RN{1}:} For MM3~\RN{1} the best-fit column density and excitation temperature values are 3.0$\times10^{15}$ cm$^{-2}$ and 220 K respectively. The uncertainty on these values is $\sim$35\% for $T_{\textrm{ex}}$ and 20\% for $N_{\textrm{s}}$. For fixed excitation temperatures down to 50 K and up to 500 K, the CH$_3$NH$_2$ column density remains within a factor of two of the best-fit value at 220 K. The value of the column density of the best-fit model does not change when the contributions from other species are included in the fit. As in the case of the MM1 region, the good agreement between the CH$_3$NH$_2$ model and data, especially around the transitions at 301.426 GHz and 301.653 GHz, makes the presence of CH$_3$NH$_2$ in this region very convincing. Due to blending with other species at the location of the CH$_3$NH$_2$ transitions at 301.248~GHz and 302.802 GHz, we consider these as tentative detections only. In the case of the transition located at 303.734 GHz, a weak line feature is present in the observed spectrum although not at the exact same location as predicted in the model spectrum. This transition is therefore also considered a tentative detection. As in the case of MM2~\RN{1}, the tentative detections are included when the model spectra are optimised.

In summary, CH$_3$NH$_2$ is securely detected towards both the MM1 and MM3 regions while the detection towards MM2 is slightly less clear. The uncertainty on the CH$_3$NH$_2$ column densities is between 15 and 20\%. Despite the local variations, the overall uniformity of CH$_3$NH$_2$ makes it likely that its origin is the same throughout the NGC 6334\RN{1} region. In addition to the data presented here, we included in Appendix \ref{app:B10} a confirmation of the presence of CH$_3$NH$_2$ in NGC 6334\RN{1} based on ALMA Band 10 observations from \citet{McGuire2018}. However, due to the difference in angular resolution and extraction location, these data probe different excitation conditions and different populations of gas and therefore cannot be compared directly with the Band 7 observations discussed above. The Band 10 spectrum and modelled CH$_3$NH$_2$ transitions shown in Fig. \ref{fig:B10}  and listed in Table \ref{tab:B10_line_summary} are therefore included as proof of the presence of CH$_3$NH$_2$ in NGC 6334\RN{1} but will not be discussed further here. A detailed analysis of the Band 10 data is presented by \citet{McGuire2018}.

\subsection{Summary of results on methanimine, methyl cyanide and formamide}
A single (hyperfine-split) transition of CH$_2$NH is covered by the data and consequently no excitation temperature can be derived for this species. In addition, the transition is blended with CH$_3$OCHO and the column density of CH$_2$NH is therefore reported as an upper limit for each of the studied regions. In contrast, a total of eleven transitions belonging to the $^{13}$C- and $^{15}$N-methyl cyanide isotopologues are detected towards NGC 6334\RN{1}. Six of these belong to $^{13}$CH$_3$CN and five to CH$_3$C$^{15}$N. Though some transitions are blended, both isotopologues are clearly detected towards all of the studied regions. The uncertainty on the derived column densities of $^{13}$CH$_3$CN and CH$_3$C$^{15}$N is up to 30\% while the uncertainty on the derived excitation temperatures is up to 65\%. In the case of MM2, the excitation temperature for CH$_3$C$^{15}$N could not be constrained and therefore the column density of this species is derived assuming $T_{\textrm{ex}}$ to be the same as for $^{13}$CH$_3$CN. As in the case of CH$_2$NH, no excitation temperature can be derived for NH$_2$CHO since only two of the 18 transitions of this species covered by the data are bright enough to be detected and these represent a very limited range of upper state energies, with a difference between the two of less than 10 K. In the case of the regions MM1~\RN{2} and MM2~\RN{1}, the features in the data at the location of the NH$_2$CHO transitions cannot be reproduced by any other species included in either the JPL or the CDMS catalogues. In contrast, the features detected towards the MM3~\RN{1} region, may be reproduced by other species and the detection of NH$_2$CHO towards this region is therefore considered tentative. The uncertainty on the column density of NH$_2$CHO towards MM1~\RN{2} and MM2~\RN{1} is less than 25\%. The full discussion of the detections of CH$_2$NH, CH$_3$CN, and NH$_2$CHO can be found in Appendix \ref{app:results}.

\section{Discussion} \label{sec:discuss}
\begin{figure*}
	\centering
	\includegraphics[width=0.78\textwidth, trim={0 0 0 0}, clip]{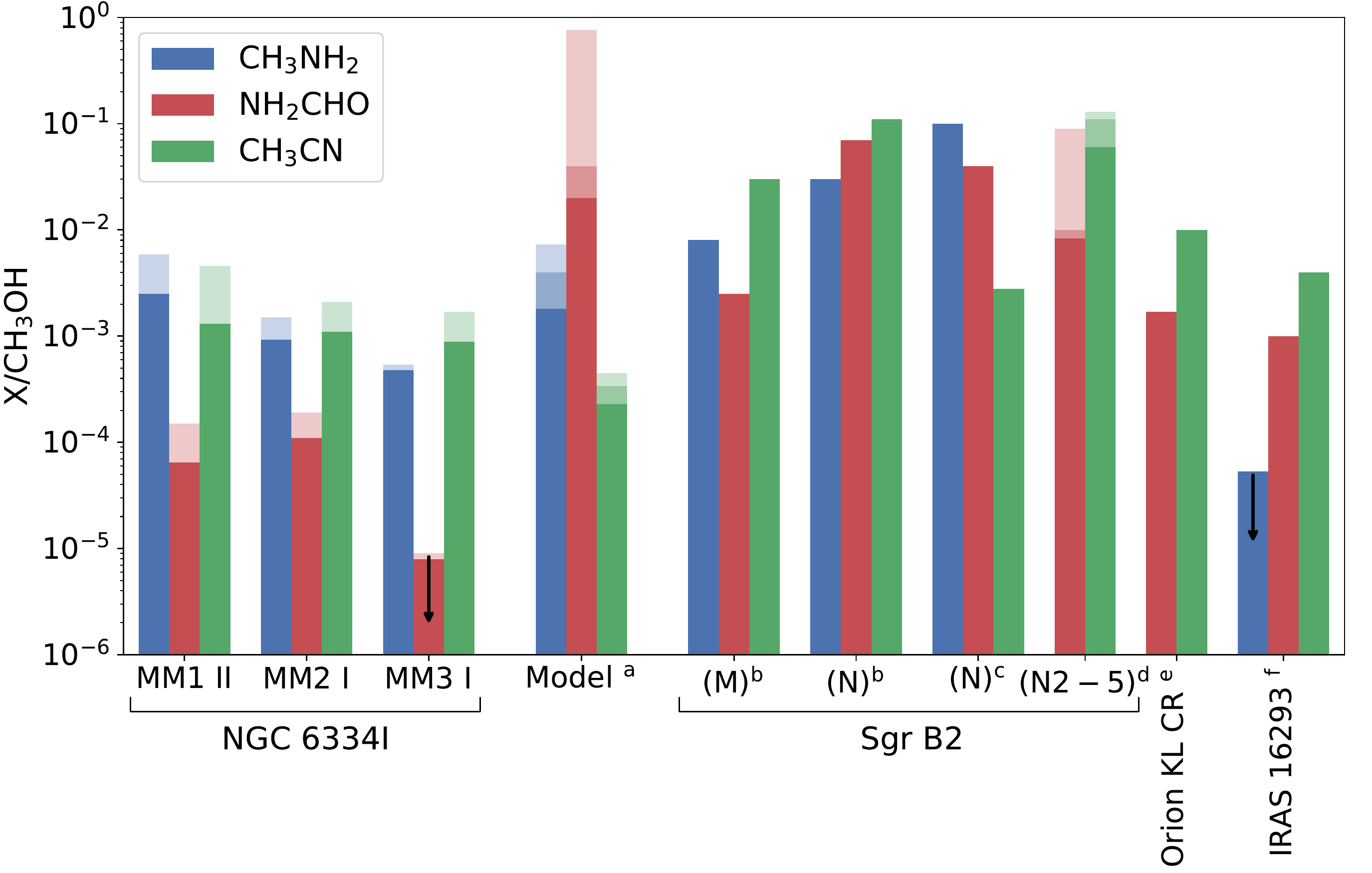} 
	\caption[]{Column density ratios of CH$_3$NH$_2$ (blue), NH$_2$CHO (red), and CH$_3$CN (green) with respect to CH$_3$OH for NGC 6334\RN{1}, model predictions and other objects. For the regions in NGC 6334\RN{1}, the shaded bars indicate the range of ratios derived using the $^{13}$C- and $^{18}$O-methanol isotopologues as base respectively. For the models, the shaded bars indicate the range of rations derived for the fast, medium and slow models respectively. For Sgr B2(N2--5), the shaded bars indicate the range of ratios derived for each of the components N2, N3, N4, and N5 (excluding the upper limit on NH$_2$CHO for N4). \textbf{References.} $^{(a)}$~\cite{Garrod2013}; $^{(b)}$~\cite{Belloche2013}; $^{(c)}$~\cite{Neill2014}; $^{(d)}$~\cite{Bonfand2017}; $^{(e)}$~\cite{Crockett2014}; $^{(f)}$~\cite{Ligterink2018}.}
	\label{fig:CH3OHratios}
\end{figure*}

\begin{figure*}
	\centering
	\includegraphics[width=0.78\textwidth, trim={0 0 0 0}, clip]{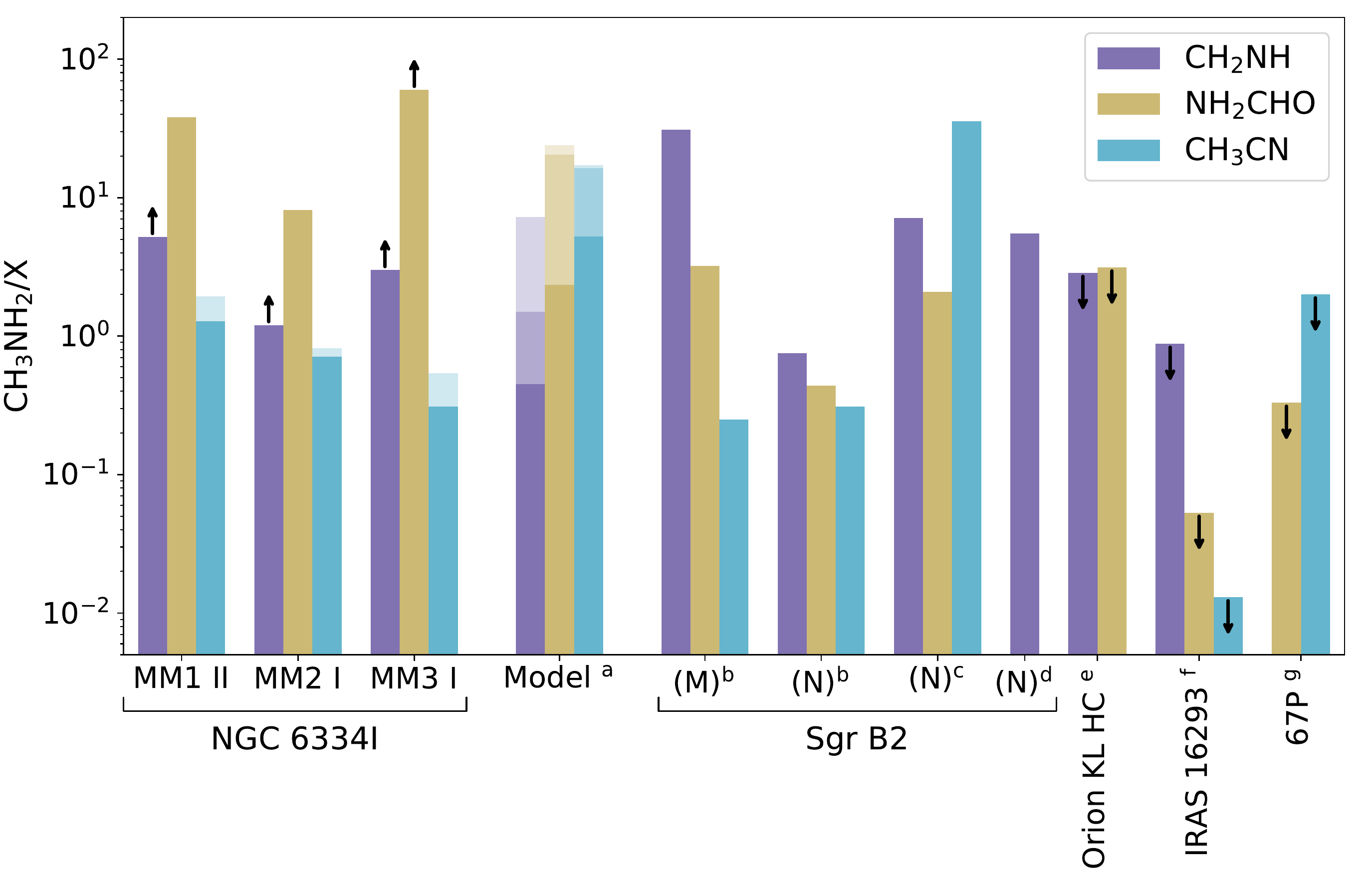} 
	\caption[]{Column density ratios of CH$_3$NH$_2$ with respect to CH$_2$NH (purple), NH$_2$CHO (gold), and CH$_3$CN (turquoise) for NGC~6334\RN{1}, model predictions and other objects. For the regions in NGC 6334\RN{1}, shaded bars indicate the range of ratios derived based on the $^{13}$CH$_3$CN and CH$_3$C$^{15}$N isotopologues. For the models, the shaded bars indicate the range of rations derived for the fast, medium and slow models respectively. \textbf{References.} $^{(a)}$~\cite{Garrod2013}; $^{(b)}$~\cite{Belloche2013}; $^{(c)}$~\cite{Neill2014}; $^{(d)}$~\cite{Halfen2013}; $^{(e)}$~\cite{Pagani2017}, Laurent Pagani, priv. comm.; $^{(f)}$~\cite{Ligterink2018}; $^{(g)}$~\cite{Goesmann2015}.}
	\label{fig:CH3NH2ratios}
\end{figure*}

In this section, the column densities and excitation temperatures discussed above will be compared with the predictions of the chemical models of \cite{Garrod2013} as well as to the values derived towards a number of other sources including the high-mass star-forming regions Sgr B2 and Orion KL, the low-mass protostar IRAS~16293--2422B and the comet 67P. In order to do this, column density ratios for each of the studied species with respect to CH$_{3}$OH are derived, these are given in Table \ref{tab:ratios_CH3OH}. CH$_3$OH is chosen as a reference because it is one of the most abundant COMs in the ISM and therefore has been studied comprehensively, also in NGC 6334\RN{1} \citep{Bogelund2018}. Secondly, in order to investigate the relation between the studied species, column density ratios of CH$_{3}$NH$_{2}$ with respect to CH$_{2}$NH, NH$_{2}$CHO, and CH$_{3}$CN are derived, these are given in Table \ref{tab:ratios_others}. Figures \ref{fig:CH3OHratios} and \ref{fig:CH3NH2ratios} summarise all ratios. In the following sections the results on CH$_{3}$NH$_{2}$ and on the other species will be discussed separately. 

\begin{table*}[]
	\begin{tiny}
		\centering
		\caption{Column density ratios with CH$_3$OH as reference}
		\label{tab:ratios_CH3OH}
		\begin{tabular}{lccccccccc}
			\toprule
			& \multicolumn{2}{c}{CH$_3$NH$_2$/CH$_3$OH} & \multicolumn{2}{c}{NH$_2$CHO/CH$_3$OH} & \multicolumn{4}{c}{CH$_3$CN/CH$_3$OH} & References \\
			& & & & & \multicolumn{2}{c}{$^{13}$CH$_3$CN} & \multicolumn{2}{c}{CH$_3$C$^{15}$N} & \\
			CH$_3$OH reference & $^{13}$C & $^{18}$O & $^{13}$C & $^{18}$O & $^{13}$C & $^{18}$O & $^{13}$C & $^{18}$O & \\
			\midrule
			MM1~\RN{2} & 5.9$\times10^{-3}$ & 2.5$\times10^{-3}$ & 1.5$\times10^{-4}$ & 6.5$\times10^{-5}$ & 4.6$\times10^{-3}$ & 2.0$\times10^{-3}$ & 3.0$\times10^{-3}$ & 1.3$\times10^{-3}$ & this work \\
			MM2~\RN{1} & 1.5$\times10^{-3}$ & 9.2$\times10^{-4}$ & 1.9$\times10^{-4}$ & 1.1$\times10^{-4}$ & 2.1$\times10^{-3}$ & 1.3$\times10^{-3}$ & 1.9$\times10^{-3}$ & 1.1$\times10^{-3}$ & this work \\
			MM3~\RN{1} & 5.4$\times10^{-4}$ & 4.8$\times10^{-4}$ & $\leq$9.0$\times10^{-6}$ & $\leq$7.9$\times10^{-6}$ & 1.0$\times10^{-3}$ & 8.9$\times10^{-4}$ & 1.7$\times10^{-3}$ & 1.5$\times10^{-3}$ & this work \\
			\midrule
			Model F & \multicolumn{2}{c}{7.3$\times10^{-3}$} & \multicolumn{2}{c}{0.04} & \multicolumn{4}{c}{4.5$\times10^{-4}$} & 1 \\
			Model M & \multicolumn{2}{c}{4.0$\times10^{-3}$} & \multicolumn{2}{c}{0.02} & \multicolumn{4}{c}{2.3$\times10^{-4}$} & 1 \\
			Model S & \multicolumn{2}{c}{1.8$\times10^{-3}$} & \multicolumn{2}{c}{0.76} & \multicolumn{4}{c}{3.4$\times10^{-4}$} & 1 \\
			\midrule
			Sgr B2(M) & \multicolumn{2}{c}{8.0$\times10^{-3}$} & \multicolumn{2}{c}{2.5$\times10^{-3}$} & \multicolumn{4}{c}{0.03} & 2 \\
			Sgr B2(N) & \multicolumn{2}{c}{0.03} & \multicolumn{2}{c}{0.07} & \multicolumn{4}{c}{0.11} & 2 \\
			Sgr B2(N) & \multicolumn{2}{c}{0.10} & \multicolumn{2}{c}{0.04} & \multicolumn{4}{c}{2.8$\times10^{-3}$} & 3 \\
			Sgr B2(N2--5)\tablefootmark{a} & \multicolumn{2}{c}{--} & \multicolumn{2}{c}{8.3$\times10^{-3}$ -- 0.09\tablefootmark{b}} & \multicolumn{4}{c}{0.06 -- 0.13} & 4 \\
			Orion KL Compact Ridge & \multicolumn{2}{c}{--} & \multicolumn{2}{c}{1.7$\times10^{-3}$} & \multicolumn{4}{c}{0.01} & 5 \\
			\midrule
			IRAS 16293--2422B & \multicolumn{2}{c}{$\leq$5.3$\times10^{-5}$} & \multicolumn{2}{c}{10$^{-3}$} & \multicolumn{4}{c}{4$\times10^{-3}$} & 6, 7, 8, 9 \\
			\bottomrule
		\end{tabular}
		\tablefoot{The uncertainty on the column density ratios derived towards NGC 6334~\RN{1} is estimated to be 40\% (see sect. \ref{sec:Method}). \tablefoottext{a}{Range of values derived for the cores N2, N3, N4, and N5.} \tablefoottext{b}{Excluding the upper limit on NH$_2$CHO for N4.}}
		\tablebib{(1)~\citet{Garrod2013}; (2) \citet{Belloche2013}; (3) \citet{Neill2014}; (4) \citet{Bonfand2017}; (5) \citet{Crockett2014}; (6) \citet{Coutens2016}; (7) \citet{Ligterink2018}; (8) \citet{Jorgensen2018}; (9) \citet{Calcutt2018}.}
	\end{tiny}
\end{table*}

\begin{table*}[]
	\centering
	\caption{CH$_3$NH$_2$ column density ratios}
	\label{tab:ratios_others}
	\begin{tabular}{lccccc}
		\toprule
		& CH$_3$NH$_2$/CH$_2$NH & CH$_3$NH$_2$/NH$_2$CHO & \multicolumn{2}{c}{CH$_3$NH$_2$/CH$_3$CN} & References \\
		CH$_3$CN reference & & & $^{13}$C & $^{15}$N & \\
		\midrule
		MM1~\RN{2} & $\geq$5.2 & 38 & 1.28 & 1.94 & this work \\
		MM2~\RN{1} & $\geq$1.2 & 8.16 & 0.71 & 0.70 & this work \\
		MM3~\RN{1} & $\geq$3.0 & $\geq$60 & 0.54 & 0.31 & this work \\
		\midrule
		Model F & 7.27 & 20.5 & \multicolumn{2}{c}{16.3} & 1 \\
		Model M & 1.5 & 24 & \multicolumn{2}{c}{17.1} & 1 \\
		Model S & 0.45 & 2.34 & \multicolumn{2}{c}{5.23} & 1 \\
		\midrule
		Sgr B2(M) & 31\tablefootmark{a} & 3.21 & \multicolumn{2}{c}{0.25} & 2 \\ 
		Sgr B2(N) & 0.75 & 0.44 & \multicolumn{2}{c}{0.31} & 2 \\ 
		Sgr B2(N) & 7.14 & 2.08 & \multicolumn{2}{c}{35.7} & 3 \\
		Sgr B2(N) & 5.49 & -- & \multicolumn{2}{c}{--} & 4 \\
		Orion KL Hot Core & $\leq$2.86 & $\leq$3.13 & \multicolumn{2}{c}{--} & 5 \\
		\midrule
		IRAS 16293--2422B & $\leq$0.88 & $\leq$0.053 & \multicolumn{2}{c}{$\leq$0.013} & 6, 7 \\
		\midrule
		Comet 67P\tablefootmark{b} & -- & $\leq$0.33 & \multicolumn{2}{c}{$\leq$2} & 8 \\
		\bottomrule
	\end{tabular}
	\tablefoot{The uncertainty on the column density ratios derived towards NGC 6334~\RN{1} is estimated to be 40\% (see sect. \ref{sec:Method}). \tablefoottext{a}{Extended CH$_2$NH emission.} \tablefoottext{b}{Listed as upper limits based on the discussion in Sect. 2.4 of \citet{Altweeg2017}.}} 
	\tablebib{(1)~\cite{Garrod2013}; (2) \cite{Belloche2013}; (3) \cite{Neill2014}; (4) \cite{Halfen2013}; (5) \cite{Pagani2017}, Laurent Pagani, priv. comm.; (6) \cite{Calcutt2018}; (7) \cite{Ligterink2018}; (8) \cite{Goesmann2015}.}
\end{table*}

\subsection{Methylamine towards NGC 6334\RN{1}}
The detection of CH$_3$NH$_2$ in the hot cores of NGC 6334\RN{1} presented here, combined with recent (tentative) detections by \citet{Pagani2017} towards Orion KL and \citet{Ohishi2017} towards a few high-mass objects, indicate that this molecule is more common and abundant than previously thought \citep[see for example the upper limits on the species presented by][]{Ligterink2015}. In this case, the `lacking' CH$_3$NH$_2$-detections are more likely explained by observational biases, for example the large partition function of CH$_{3}$NH$_{2}$ resulting in relatively weaker transitions of this species as compared with, for example, NH$_2$CHO, rather than actual chemical variations between objects.  

Within the regions of NGC 6334\RN{1}, the CH$_{3}$NH$_{2}$ abundance is fairly uniform and column density ratios with respect to CH$_{3}$OH and CH$_{3}$CN show variations within factors of four and two between regions MM1 and MM2 and up to an order of magnitude between regions MM1 and MM3. The variation over the column density ratios derived using the $^{13}$C- and $^{18}$O-methanol iosotopologues as a reference vary with a factor of three, while the ratios derived based on the $^{13}$C- and $^{15}$N-methyl cyanide isotopologues vary with a factor of two. In the case of the CH$_{3}$NH$_{2}$ to NH$_{2}$CHO ratio, the variation is a factor of seven if all three regions are considered and less than a factor of five between regions MM1~\RN{2} and MM2~\RN{1}. This is due to the relatively low column density of NH$_{2}$CHO in MM3~\RN{1} as compared with regions MM1~\RN{2} and MM2~\RN{1}. Similarly, the variation of the CH$_3$NH$_2$ to CH$_2$NH column density ratio over the three regions is within a factor of four, though the single CH$_2$NH line covered by the data means that these ratios should be seen as lower limits. 

Although the variations in the column density of CH$_3$NH$_2$ over the studied regions are similar to those of CH$_3$OH and CH$_3$CN, the CH$_3$NH$_2$ excitation temperatures are higher than for any of the other species. This trend is most pronounced in the case of MM1. A relatively higher excitation temperature of CH$_3$NH$_2$ compared with other species is consistent with the findings of \citet{Halfen2013}.

\subsection{Methylamine towards other objects}
Compared with the CH$_{3}$NH$_{2}$ to CH$_{3}$OH ratios derived by \cite{Belloche2013} and \cite{Neill2014} towards Sgr B2 (M) and (N), the values inferred for the regions in NGC 6334\RN{1} are lower by up to two orders of magnitude, though the value derived for Sgr B2 (M) is only higher by a factor of up to three when compared with the value derived for MM1. For the CH$_{3}$NH$_{2}$ to NH$_2$CHO, CH$_3$CN, and CH$_2$NH ratios the picture is less clear; while the CH$_3$NH$_2$/NH$_2$CHO values derived towards Sgr~B2 are all about an order of magnitude lower than those derived towards NGC~6334\RN{1}, the CH$_3$NH$_2$/CH$_3$CN value derived by \citet{Neill2014} is higher by more than an order of magnitude while the values derived by \citet{Belloche2013} are lower by up to a factor of six. In the case of CH$_3$NH$_2$/CH$_2$NH, all but one of the values towards Sgr B2 are higher than the lower limits derived towards NGC 6334\RN{1}. Because of these large variations it is difficult to make strong statements on the overall CH$_{3}$NH$_{2}$ distribution within the Sgr B2 region since chemical variations in the reference species are just as likely the source of the varying ratios. Also, due to the large distance to Sgr B2 ($\sim$8 kpc) and the fact that \citet{Belloche2013} and \citet{Neill2014} use single dish data, from the IRAM 30 m telescope and \textit{Herschel} Space Observatory respectively, the observations may be biased towards large scale structures and particularly the effects of beam dilution should be considered since these studies probe spacial scales of the order of $\sim$0.5-1 pc ($\sim$2$\times$10$^5$ au) as compared with $\sim$1300 au in the case of the regions in NGC~6334\RN{1}. 

In contrast to the studies of Sgr B2, the ALMA observations towards the Orion KL Hot Core region reported by \citet{Pagani2017} make for a more direct comparison with the observations towards NGC~6334\RN{1}, since the Orion KL region is probed at spacial scales of $\sim$660 au. Though not firmly detected, the upper limits on the CH$_{3}$NH$_{2}$ to CH$_2$NH or NH$_2$CHO ratios hint that CH$_{3}$NH$_{2}$ is less abundant in the Orion KL hot core as compared with NGC 6334\RN{1} or, alternatively, that NH$_2$CHO and CH$_2$NH are more abundant. Unfortunately the extended CH$_3$OH emission towards Orion KL could not be evaluated due to missing zero-spacing data. Without CH$_3$OH as a reference it is difficult to distinguish between the low-CH$_{3}$NH$_{2}$ and high-NH$_2$CHO or CH$_2$NH scenarios. In addition, the Orion KL data show that CH$_{3}$NH$_{2}$ is not associated with either NH$_{2}$CHO nor CH$_{2}$NH. This is based on the $v_{\rm LSR}$ which is 4.3 km s$^{-1}$ for CH$_3$NH$_2$ but 5.5 km s$^{-1}$ for NH$_{2}$CHO and CH$_{2}$NH. In NGC 6334\RN{1} such a mismatch between velocity of different species is not observed toward either of the studied regions. Finally, as in the case of NGC 6334\RN{1}, the excitation temperature is higher for CH$_3$NH$_2$ than for NH$_{2}$CHO and CH$_{2}$NH with values of 280, 200, and 150 K respectively. 

The lowest CH$_{3}$NH$_{2}$ ratios are observed towards the low-mass protostar IRAS~16293--2422B, an analogue to the young Sun, where a deep upper limit on the column density of CH$_3$NH$_2$ was inferred by \citet{Ligterink2018}, based on the ALMA PILS survey \citep[see][for full PILS overview]{Jorgensen2016} probing spacial scales of $\sim$60 au. This upper limit results in ratios with respect to CH$_3$OH, NH$_2$CHO, and CH$_3$CN which are all lower by one to two orders of magnitude when compared with the lowest ratios derived towards NGC 6334\RN{1}. The smallest variation between NGC 6334\RN{1} and IRAS~16293--2422B is seen in the CH$_{3}$NH$_{2}$ to CH$_2$NH ratio where the value derived for IRAS~16293--2422B is within the uncertainty of value derived for the MM2~\RN{1} region but lower by up to a factor of six compared with the regions MM1~\RN{2} and MM3~\RN{1}. These differences in ratios hint that the formation of CH$_{3}$NH$_{2}$ in the high-mass hot cores of NGC 6334\RN{1} differ from the formation of CH$_{3}$NH$_{2}$ in the low-mass IRAS~16293--2422B protostar. An explanation for this difference could be the dust grain temperature. Based on the low levels of CH$_3$OH deuteration in NGC 6334\RN{1}, \citet{Bogelund2018} determine a relatively warm dust grain temperature of $\sim$30~K during the time of CH$_{3}$OH formation. In contrast, the dust grains in the cloud from which the IRAS~16293--2422 protobinary system formed are thought to have been much cooler, with temperatures below 20 K \citep{Jorgensen2016}. At high grain temperatures the solid-state formation of CH$_{3}$NH$_{2}$ via CH$_{3}$ + NH$_{2}$ could be enhanced, due to increased mobility of the radicals or the loss of H-atoms, which at lower temperatures would hydrogenate these radicals to form the neutral species CH$_4$ and NH$_3$. 

Additional indications for a grain surface formation route are found in the chemical models presented by \citet{Garrod2013}. These models evaluate the chemical evolution of high-mass hot cores as these evolve through infall and warm-up phases. The physical model adopted by \citet{Garrod2013} consists of a collapse phase followed by a gradual warm-up of the gas and dust. For the warm-up phase, three timescales are adopted: a `fast' scale reaching 200~K in 5$\times$10$^4$~yr, a `medium' scale reaching 200~K in 2$\times$10$^5$~yr, and a `slow' scale reaching 200~K in 1$\times$10$^6$~yr. Listed in Tables~\ref{tab:ratios_CH3OH} and \ref{tab:ratios_others} are the predicted peak gas-phase abundance ratios for each of these models. In the models, CH$_{3}$NH$_{2}$ is formed predominantly via CH$_{3}$ and NH$_{2}$ radical recombination reactions on the grain surface. Since the predicted CH$_{3}$NH$_{2}$ ratios are quite similar to the ratios derived for the regions in NGC 6334\RN{1}, and for most species agree within a factor of five, a solid state formation pathway for CH$_{3}$NH$_{2}$ seems likely. However, since the models are not optimised to the physical conditions of the hot cores of NGC 6334~\RN{1} but rather general conditions found in hot cores, the comparison between observed and modelled column density ratios should only be considered as indicative of trends.

\subsection{Comparison with comet 67P} \label{sebsec:comet67p}
In an effort to understand how the life we know on Earth today has come to be, the chemical composition of the Solar Nebular must be examined. The most pristine record of this composition is believed to be locked up in comets. \citet{Goesmann2015} report the first in situ analysis of organic molecules on the surface of comet 67P. Based on the measurements of the COSAC instrument aboard \textit{Rosetta}'s \textit{Philae} lander, \citet{Goesmann2015} derive CH$_3$NH$_2$ to NH$_2$CHO and CH$_3$CN ratios which are lower by one to two orders of magnitude for CH$_3$NH$_2$/NH$_2$CHO and higher by up to a factor of six for CH$_3$NH$_2$/CH$_3$CN, as compared with the values derived for NGC 6334\RN{1}. To improve counting statistics, \citet{Goesmann2015} binned the COSAC data in bins around integer mass numbers, thereby effectively reducing the mass resolution, before identifying and deriving abundances of the detected species. However, after reanalysing the unbinned COSAC data, and using higher resolution measurements from the ROSINA mass spectrometer, aboard the \textit{Rosetta} orbiter, as a proxy for the near-surface cometary material, \citet{Altweeg2017} conclude that a revision of the list of molecules and derived abundances reported by \citet{Goesmann2015} is needed. Specifically, the contributions from CH$_3$NH$_2$, NH$_2$CHO, and CH$_3$CN to the signal in the COSAC data are likely to be significantly smaller than originally reported by \citet{Goesmann2015}. Therefore, the CH$_3$NH$_2$ ratios for comet 67P are listed in this work as upper limits (following the discussion in Sect. 2.4 of \citet{Altweeg2017}). The ratios derived for the comet are consistent with the values derived for the low-mass protostar IRAS~16293--2422B.    

\subsection{Other N-bearing species}
For the NH$_2$CHO and CH$_3$CN to CH$_3$OH ratios, the variations derived for each of the NGC 6334\RN{1} regions are small and within factors of between two and four (excluding the upper limit on NH$_2$CHO for region MM3~\RN{1} which is about an order of magnitude lower than the values for MM1~\RN{2} and MM2~\RN{1}). Compared with the hot core model predictions of \cite{Garrod2013}, NH$_2$CHO/CH$_3$OH is over-predicted by orders of magnitude, while CH$_{3}$CN/CH$_{3}$OH, as is the case for CH$_3$NH$_2$/CH$_3$OH, shows fairly good agreement with the numbers derived for NGC 6334\RN{1}.

For Sgr B2, the NH$_2$CHO and CH$_3$CN ratios with respect to CH$_3$OH show the same trends as CH$_3$NH$_2$/CH$_3$OH, and are generally one to two orders of magnitude higher than the values derived for NGC 6334\RN{1}, though, as in the case of the CH$_3$NH$_2$ ratios, observations may suffer from beam dilution effects or underestimated CH$_3$OH values since only the main CH$_3$OH-isotope, which may be optically thick, is detected. Although CH$_3$NH$_2$ is not included in their study, the ratios derived for NH$_2$CHO and CH$_3$CN by \cite{Bonfand2017}, using ALMA observations which probe scales of $\sim$0.06 pc ($\sim$13300 au), indicate that the higher NH$_2$CHO and CH$_3$CN to CH$_3$OH ratios reported by \citet{Belloche2013} and \citet{Neill2014}, are true and not artefacts of beam dilution or opacity effects. This implies that the chemical inventory of Sgr B2 is richer in complex nitrogen-bearing species than that of NGC 6334\RN{1}, in agreement with the high temperatures and complexity characterising the Galactic central region. That the NGC 6334I region is relatively poor in N-bearing species is also in agreement with the findings of \citet{Suzuki2018} who investigate the correlation between O- and N-bearing species in a sample of eight hot cores and find that the former species are more abundant than the latter in this region.

For the Orion KL Compact Ridge, \citet{Crockett2014} use observations from \textit{Herschel} to derive NH$_2$CHO/CH$_3$OH and CH$_3$CN/CH$_3$OH values which are generally lower than those derived for Sgr B2 but higher by at least an order of magnitude as compared with NGC 6334\RN{1}.

Lastly, the ALMA observations towards the low-mass protostar IRAS 16293--2422B, indicate similar CH$_{3}$CN/CH$_{3}$OH values as compared with the regions in NGC 6334\RN{1}, while the values for NH$_2$CHO/CH$_3$OH are higher for IRAS 16293--2422B by about an order of magnitude as compared with the values for the regions in NGC 6334\RN{1}. The generally similar CH$_3$CN and NH$_2$CHO to CH$_3$OH ratios between NGC 6334\RN{1} and IRAS 16293--2422B indicate that the overall lower CH$_3$NH$_2$ ratios derived towards IRAS 16293--2422B reflect an actual difference in chemical composition between the two regions. As discussed above, this difference in CH$_3$NH$_2$ abundance may reflect a difference in grain temperature during the time when the species was formed. With the sensitivity and resolution provided by ALMA, continued studies of this and related species will broaden our understanding of the inventory of pre-biotic species in both high- and low-mass sources and help evaluate the degree to which CH$_3$NH$_2$ chemistry depends on the grain temperature.

\section{Summary} \label{sec:conclusion}
In this work, we present the first detection of CH$_3$NH$_2$ towards NGC 6334\RN{1} and derive the column density of the species in the hot cores MM1, MM2, and MM3. Transitions of CH$_2$NH, NH$_2$CHO, $^{13}$CH$_3$CN, and CH$_3$C$^{15}$N are also studied and their column densities inferred. Assuming LTE and excitation temperatures in the range 70~--~340~K, each species is modelled separately and then summed to obtain a full spectrum for each of the studied regions. Based on the good agreement between the CH$_3$NH$_2$ column density ratios predicted by the hot core models of \citet{Garrod2013} and the values derived for the regions in NGC~6334\RN{1}, the formation of CH$_3$NH$_2$ is more likely to proceed via radical recombination reactions on grain surfaces than via gas-phase reactions.

The detection of CH$_3$NH$_2$ towards NGC 6334\RN{1} reported here and recent (tentative) detections towards the high-mass star-forming regions in Orion KL and G10.47+0.03 by \citet{Pagani2017} and \citet{Ohishi2017} respectively, also indicate that the species is not as uncommon in the ISM as was previously thought. This implies that future high-sensitivity, high-resolution searches for the species are likely to yield additional detections of the formerly so elusive molecule. In this case, observations carried out towards both high- and low-mass objects, will help assess the dependency of CH$_3$NH$_2$-grain formation efficiency on the dust grain temperature of individual regions. 

\begin{acknowledgements} 
We thank the anonymous referee for a careful evaluation and many useful comments that helped us clarify our manuscript.	A special thanks to L. Pagani for insights into the complex structure and chemistry of Orion KL and providing column density estimates for NH$_2$CHO and CH$_2$NH. We also thank C. Brogan and T. Hunter for assistance in reducing and analysing the Band 10 data. This paper makes use of the following ALMA data: ADS/JAO.ALMA\#2015.1.00150.S and \#2017.1.00717.S. ALMA is a partnership of ESO (representing its member states), NSF (USA) and NINS (Japan), together with NRC (Canada) and NSC and ASIAA (Taiwan) and KASI (Republic of Korea), in cooperation with the Republic of Chile. The Joint ALMA Observatory is operated by ESO, AUI/NRAO and NAOJ.
This work is based on analysis carried out with the CASSIS software and JPL: http://spec.jpl.nasa.gov/ and CDMS: http://www.ph1.uni-koeln.de/cdms/ spectroscopic databases. CASSIS has been developed by IRAP-UPS/CNRS (http://cassis.irap.omp.eu). 
\end{acknowledgements}

\bibliographystyle{aa} 
\bibliography{BibTex/NGC6334I_CH3NH2} 

\appendix
\section{Model grids}
\begin{table}[h]
	\centering
	\caption{Overview of model grids}
	\label{tab:model_grid}
	\begin{tabular}{lccc}
		\toprule
		Species & \multicolumn{3}{c}{$N_{\textrm{s}}$ range [cm$^{-2}$]} \\
		\cline{2-4}
		& MM1~\RN{2} & MM2~\RN{1} & MM3~\RN{1} \\ 
		\midrule	
		CH$_2$NH & 10$^{16}$ -- 10$^{17}$ & 10$^{16}$ -- 10$^{17}$ & 5$\times$10$^{14}$ -- 5$\times$10$^{15}$ \\
		CH$_3$NH$_2$ & 5$\times$10$^{16}$ -- 5$\times$10$^{17}$ & 3$\times$10$^{16}$ -- 3$\times$10$^{17}$ & 10$^{15}$ -- 10$^{16}$ \\
		$^{13}$CH$_3$CN & 10$^{15}$ -- 10$^{16}$ & 5$\times$10$^{14}$ -- 5$\times$10$^{15}$ & 5$\times$10$^{13}$ -- 5$\times$10$^{14}$ \\
		CH$_3$C$^{15}$N & 10$^{14}$ -- 10$^{15}$ & 10$^{14}$ -- 10$^{15}$ & 5$\times$10$^{12}$ -- 5$\times$10$^{13}$ \\
		NH$_2$CHO & 5$\times$10$^{15}$ -- 5$\times$10$^{16}$ & 5$\times$10$^{15}$ -- 5$\times$10$^{16}$ & 10$^{13}$ -- 10$^{14}$ \\
		NH$_2^{13}$CHO & 5$\times$10$^{14}$ -- 5$\times$10$^{15}$ & 10$^{14}$ -- 10$^{15}$ & -- \\
		\bottomrule
	\end{tabular}
	\tablefoot{All grids have $T_{\textrm{ex}}$ spanning 10 -- 500 K in steps of 10 K and $N_{\textrm{s}}$ sampled by 20 logarithmically spaced steps.}
\end{table}

\clearpage
\onecolumn
\section{Integrated line intensities} \label{app:Integrated_lines}
This appendix lists the integrated intensities of the best-fit model for each species and region, 
along with the integrated intensity, FWHM and $v_{\textrm{LSR}}$ of a gaussian profile fitted to selected line features in the observed spectra. However, due to the high line density in the observed spectra, the majority of the listed transitions are blended. Therefore, care should be taken when interpreting the integrated intensities of the observed transitions since these fits in the majority of cases cover blended features which cannot be disentangled and therefore will included the contributions from other (unknown) species. 

\begin{center}
	\begin{longtable}{cccccccc}
			\caption{Integrated intensities of spectral line features} \\
			\toprule
			\label{tab:integrated_lines}
			& & & & & \multicolumn{3}{c}{Fit to observed spectrum\tablefootmark{a}} \\
			\cmidrule{6-8}
			\multicolumn{2}{c}{Transition} & & Region & $I_\textrm{model}$\tablefootmark{b} & $I_{\textrm{gauss}}$ & FWHM$_{\textrm{gauss}}$ & $v_{\textrm{LSR,gauss}}$ \\
			\cmidrule{1-2}
			$[\textrm{QN}]_{\textrm{up}}$\tablefootmark{c} & $[\textrm{QN}]_{\textrm{low}}$\tablefootmark{c} & &  & [K km s$^{-1}$] & [K km s$^{-1}$] & [km s$^{-1}$] & [km s$^{-1}$] \\
			\midrule
			\endfirsthead
			\multicolumn{8}{c}{\textit{Continued from previous page}}\\
			\hline
			& & & & & \multicolumn{3}{c}{Fit to observed spectrum\tablefootmark{a}} \\
			\cmidrule{6-8}
			\multicolumn{2}{c}{Transition} & & Region & $I_\textrm{model}$\tablefootmark{b} & $I_{\textrm{gauss}}$ & FWHM$_{\textrm{gauss}}$ & $v_{\textrm{LSR,gauss}}$ \\
			\cmidrule{1-2}
			$[\textrm{QN}]_{\textrm{up}}$\tablefootmark{c} & $[\textrm{QN}]_{\textrm{low}}$\tablefootmark{c} & &  & [K km s$^{-1}$] & [K km s$^{-1}$] & [km s$^{-1}$] & [km s$^{-1}$] \\
			\midrule  
			\endhead
			\midrule
			\multicolumn{8}{c}{\textit{Continued on next page}}\\
			\endfoot
			\bottomrule
			\endlastfoot
			\multicolumn{8}{c}{$\rm{CH_2NH}$} \vspace{0.2cm} \\
			
			15$_{2,13}$ 14 & 14$_{3,12}$ 13 & \multirow{3}{*}{\scalerel*[1ex]{\}}{\rule[-2.8ex]{1ex}{7.ex}}} & MM1\tablefootmark{*} & 46.2 & 37.8 & 2.7 $\pm$ 0.8 & -8.2 $\pm$ 0.3 \\
			15$_{2,13}$ 16 & 14$_{3,12}$ 15 & & MM2\tablefootmark{*} & 36.8 & 24.1 & 2.4 $\pm$ 0.3 & -10.4 $\pm$ 0.1\phantom{*} \\
			15$_{2,13}$ 15 & 14$_{3,12}$ 14 & & MM3\phantom{*} & 0.5 & -- & -- & -- \\
			
			& & & & & & & \\
			\multicolumn{8}{c}{$\rm{CH_3NH_2}$\tablefootmark{d}} \vspace{0.2cm} \\
			
			16 2 A2 15 & 15 3 A1 14 & \multirow{3}{*}{\scalerel*[1ex]{\}}{\rule[-2.8ex]{1ex}{7.ex}}} & MM1\phantom{*} & 18.3 & -- & -- & -- \\ 
			16 2 A2 17 & 15 3 A1 16 & & MM2\phantom{*} & 4.8 & -- & -- & -- \\
			16 2 A2 16 & 15 3 A1 16 & & MM3\phantom{*} & 0.2 & -- & -- & -- \\
			& & & & & & & \vspace{-0.1cm} \\
			13 2 B2 13 & 13 1 B1 13 & \multirow{3}{*}{\scalerel*[1ex]{\}}{\rule[-2.8ex]{1ex}{7.ex}}} & MM1\tablefootmark{*} & 106.0 & 123.6 & 4.8 $\pm$ 0.8 & -7.9 $\pm$ 0.2 \\
			13 2 B2 14 & 13 1 B1 14 & & MM2\tablefootmark{*} & 32.5 & 74.0 & 3.6 $\pm$ 0.3 & -11.0 $\pm$ 0.1\phantom{*} \\
			13 2 B2 13 & 13 1 B1 12 & & MM3\phantom{*} & 1.7 & 2.0 & 3.2 $\pm$ 0.6 & -10.9 $\pm$ 0.2\phantom{*} \\
			
			9 0 B2 8\phantom{*} & 8 1 B1 7 & \multirow{3}{*}{\scalerel*[1ex]{\}}{\rule[-2.8ex]{1ex}{7.ex}}} & MM1\phantom{*} & 115.3 & -- & -- & -- \\ 
			9 0 B2 10 & 8 1 B1 9 & & MM2\phantom{*} & 41.5 & -- & -- & --\\
			9 0 B2 9\phantom{*} & 8 1 B1 8 & & MM3\tablefootmark{*} & 2.2 & 2.4 & 4.4 $\pm$ 1.1 & -10.7 $\pm$ 0.4\phantom{*} \\
			& & & & & & & \vspace{-0.1cm} \\
			16 7 B1 16 & 17 6 B2 17 & \multirow{6}{*}{\scalerel*[1ex]{\}}{\rule[-2.8ex]{1ex}{14.ex}}} & \multirow{4}{*}{MM1\phantom{*}} & \multirow{4}{*}{36.6} & \multirow{4}{*}{--} & \multirow{4}{*}{--} & \multirow{4}{*}{--}\\ 
			16 7 B2 16 & 17 6 B1 17 & & \multirow{4}{*}{MM2\phantom{*}} & \multirow{4}{*}{7.6} & \multirow{4}{*}{--} & \multirow{4}{*}{--} & \multirow{4}{*}{--} \\
			16 7 B1 17 & 17 6 B2 18 & & \multirow{4}{*}{MM3\phantom{*}} & \multirow{4}{*}{0.4} & \multirow{4}{*}{--} & \multirow{4}{*}{--} & \multirow{4}{*}{--} \\
			16 7 B2 17 & 17 6 B1 18 & & & & & & \\
			16 7 B1 15 & 17 6 B2 16 & & & & & & \\
			16 7 B2 15 & 17 6 B1 16 & & & & & & \\
			& & & & & & & \vspace{-0.1cm} \\
			9 0 E2+1 8\phantom{*} & 8 1 E2+1 7 & \multirow{3}{*}{\scalerel*[1ex]{\}}{\rule[-2.8ex]{1ex}{7.ex}}} & MM1\phantom{*} & 38.8 & -- & -- & -- \\
			9 0 E2+1 10 & 8 1 E2+1 9 & & MM2\phantom{*} & 13.7 & -- & -- & --\\
			9 0 E2+1 9\phantom{*} & 8 1 E2+1 8 & & MM3\phantom{*} & 0.7 & -- & -- & -- \\
			
			& & & & & & & \\
			\multicolumn{8}{c}{$\rm{^{13}CH_3CN}$} \vspace{0.2cm} \\
			
			& & & MM1\tablefootmark{*}  & 27.5 & 36.9 & 3.6 $\pm$ 0.6 & -6.8 $\pm$ 0.3 \\
			17$_{5}$ & 16$_{5}$ & & MM2\tablefootmark{*}  & 10.8 & 32.2 & 2.8 $\pm$ 0.2 & -8.0 $\pm$ 0.1 \\
			& & & MM3\phantom{*} & 0.9 & 1.1 & 2.4 $\pm$ 0.3 & -9.2 $\pm$ 0.1 \\
			& & & & & & & \vspace{-0.1cm} \\
			& & & MM1\tablefootmark{*} & 49.0 & 38.1 & 2.6 $\pm$ 0.4 & -6.5 $\pm$ 0.2 \\
			17$_{4}$ & 16$_{4}$ & & MM2\tablefootmark{*}  & 24.2 & 50.4 & 3.3 $\pm$ 0.3 & -8.6 $\pm$ 0.1 \\
			& & & MM3\phantom{*} & 1.9 & 1.9 & 2.5 $\pm$ 0.2 & -8.8 $\pm$ 0.1 \\
			& & & & & & & \vspace{-0.1cm} \\
			& & & MM1\phantom{*} & 132.1 & 108.1 & 3.5 $\pm$ 0.4 & -6.4 $\pm$ 0.2 \\
			17$_{3}$ & 16$_{3}$ & & MM2\phantom{*} & 80.9 & 86.2 & 3.4 $\pm$ 0.2 & -8.3 $\pm$ 0.1 \\
			& & & MM3\phantom{*} & 6.7 & 7.1 & 2.3 $\pm$ 0.2 & -8.7 $\pm$ 0.1 \\
			& & & & & & & \vspace{-0.1cm} \\
			& & & MM1\tablefootmark{*}  & 100.4 & 124.5 & 4.4 $\pm$ 0.7 & -6.7 $\pm$ 0.2 \\
			17$_{2}$ & 16$_{2}$ & & MM2\tablefootmark{*}  & 66.9 & 72.9 & 3.2 $\pm$ 0.2 & -8.3 $\pm$ 0.1 \\
			& & & MM3\tablefootmark{*}  & 5.1 & 5.2 & 2.4 $\pm$ 0.6 & -8.6 $\pm$ 0.2 \\
			& & & & & & & \vspace{-0.1cm} \\
			\newpage
			& & & MM1\tablefootmark{*}  & 113.1 & 212.7 & 4.3 $\pm$ 0.3 & -7.5 $\pm$ 0.1 \\
			17$_{1}$ & 16$_{1}$ & & MM2\tablefootmark{*}  & 82.5 & 78.0 & 3.1 $\pm$ 0.3 & -8.4 $\pm$ 0.1 \\
			& & & MM3\tablefootmark{*}  & 6.6 & 7.5 & 2.3 $\pm$ 0.3 & -8.7 $\pm$ 0.1 \\
			& & & & & & & \vspace{-0.1cm} \\
			& & & MM1\phantom{*} & 128.5 & -- & -- & -- \\
			17$_{0}$ & 16$_{0}$ & & MM2\phantom{*} & 91.8 & -- & -- & -- \\
			& & & MM3\phantom{*} & 7.5 & -- & -- & -- \\
			
			& & & & & & & \\
			\multicolumn{8}{c}{$\rm{CH_3C^{15}N}$} \vspace{0.2cm} \\
			
			& & & MM1\phantom{*} & 8.3 & -- & -- & -- \\
			17$_{4}$ & 16$_{4}$ & & MM2\phantom{*} & 3.3 & -- & -- & -- \\
			& & & MM3\phantom{*} & 0.3 & -- & -- & -- \\
			& & & & & & & \vspace{-0.1cm} \\
			& & & MM1\phantom{*} & 26.0 & 23.7 & 2.5 $\pm$ 0.5 & -6.3 $\pm$ 0.2 \\
			17$_{1}$ & 16$_{1}$ & & MM2\tablefootmark{*}  & 12.3 & 32.56 & 2.4 $\pm$ 0.2 & -7.9 $\pm$ 0.1 \\
			& & & MM3\phantom{*} & 1.4 & 1.31 & 2.7 $\pm$ 0.6 & -8.9 $\pm$ 0.2 \\
			& & & & & & & \vspace{-0.1cm} \\
			& & & MM1\tablefootmark{*} & 18.6 & 20.2 & 3.2 $\pm$ 0.5 & -6.5 $\pm$ 0.2 \\
			17$_{2}$ & 16$_{2}$ & & MM2\tablefootmark{*}  & 9.8 & 23.0 & 3.8 $\pm$ 0.5 & -7.7 $\pm$ 0.2 \\
			& & & MM3\phantom{*} & 1.2 & 0.65 & 1.6 $\pm$ 0.4 & -8.5 $\pm$ 0.2 \\
			& & & & & & & \vspace{-0.1cm} \\
			& & & MM1\phantom{*}& 24.7 & -- & -- & -- \\
			17$_{1}$ & 16$_{1}$ & & MM2\phantom{*} & 13.0 & -- & -- & -- \\
			& & & MM3\phantom{*} & 1.6 & -- & -- & -- \\
			& & & & & & & \vspace{-0.1cm} \\
			& & & MM1\phantom{*} & 26.7 & -- & -- & -- \\
			17$_{1}$ & 16$_{1}$ & & MM2\phantom{*} & 16.5 & -- & -- & -- \\
			& & & MM3\phantom{*} & 1.8 & -- & -- & -- \\
			
			& & & & & & & \\
			\multicolumn{8}{c}{$\rm{NH_2CHO}$} \vspace{0.2cm} \\
			
			& & & MM1\phantom{*} & 135.1 & -- & -- & -- \\
			15$_{1,15}$ & 14$_{1,14}$ & & MM2\phantom{*} & 172.0 & -- & -- & -- \\
			& & & MM3\phantom{*} & 1.6 & -- & -- & -- \\
			& & & & & & & \vspace{-0.1cm} \\
			& & & MM1\phantom{*} & 130.7 & -- & -- & -- \\
			14$_{1,13}$ & 13$_{1,12}$ & & MM2\phantom{*} & 168.3 & -- & -- & -- \\
			& & & MM3\phantom{*} & 1.6 & -- & -- & -- \\
			
			& & & & & & & \\
			\multicolumn{8}{c}{$\rm{NH_2^{13}CHO}$} \vspace{0.2cm} \\
			
			& & & MM1\phantom{*} & 41.4 & -- & -- & -- \\
			15$_{1,15}$ & 14$_{1,14}$ & & MM2\phantom{*} & 13.2 & -- & -- & -- \\
			& & & MM3\phantom{*} & -- & -- & -- & -- \\
			& & & & & & & \vspace{-0.1cm} \\
			& & & MM1\phantom{*} & 40.1 & -- & -- & -- \\
			14$_{1,13}$ & 13$_{1,12}$ & & MM2\phantom{*} & 12.9 & -- & -- & -- \\
			& & & MM3\phantom{*} & -- & -- & -- & -- \\
			\bottomrule
		\end{longtable}
		\tablefoot{\tablefoottext{*}{Fit to blended feature.} \tablefoottext{a}{Gaussian fit to the observed line features. Listed values are the integrated intensity, FWHM and $v_{\textrm{LSR}}$ of the fitted gaussian profile, including $1\sigma$ uncertainties.} \tablefoottext{b}{Integrated intensity of the best-fit model for each region and spectral feature.} \tablefoottext{c}{Quantum numbers for CH$_2$NH are J$_{\textrm{Ka,Kc}}$~F. Quantum numbers for CH$_3$NH$_2$ are J $\textrm{K}_{\textrm{a}}$ $\Gamma$ F, following the notation of \citet{Motiyenko2014}. Quantum numbers for $^{13}$CH$_3$CN and CH$_3$C$^{15}$N are J$_{\textrm{K}}$. Quantum numbers for NH$_2$CHO and NH$_2^{13}$CHO are J$_{\textrm{Ka,Kc}}$.} \tablefoottext{d}{Only lines with $A_{\textrm{ij}}>10^{-6} s^{-1}$ are listed.}}
\end{center}

\clearpage
\section{ALMA Band 10 spectrum of methylamine} \label{app:B10}
\begin{figure}[h]
	\centering
	\includegraphics[width=1.\textwidth, trim={0 0 0 0}, clip]{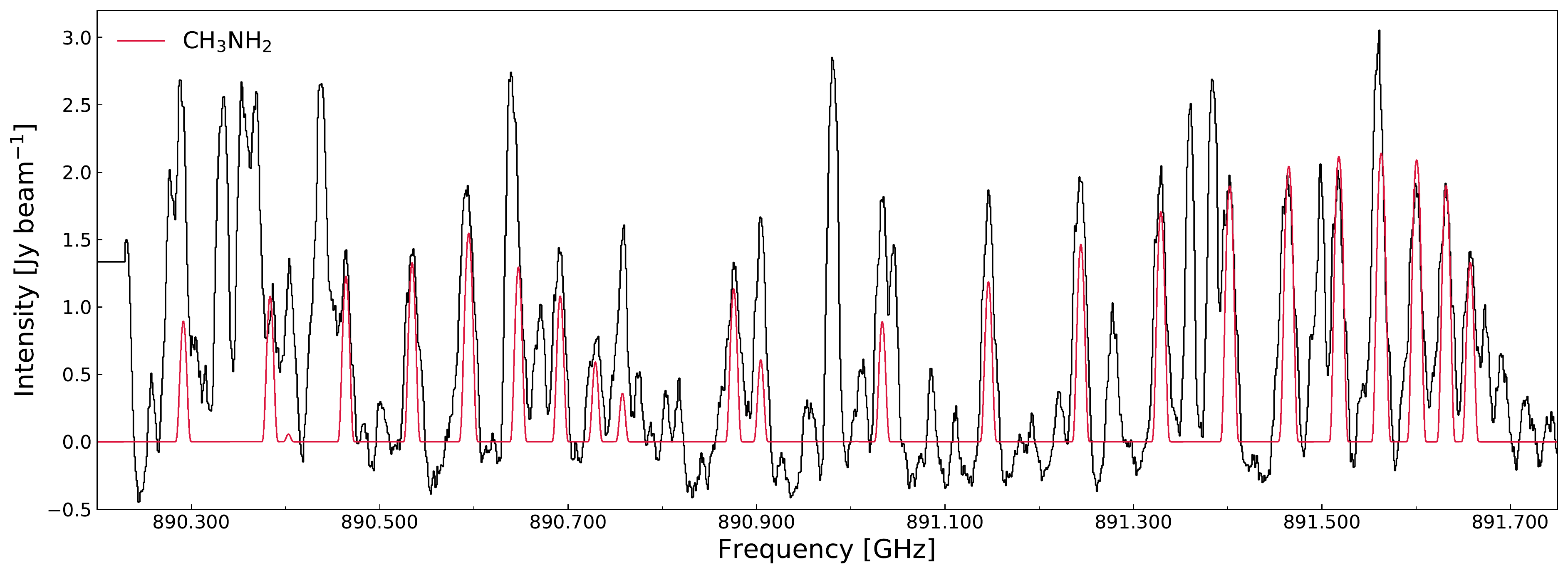} 
	\caption[]{CH$_3$NH$_2$ transitions detected towards NGC 6334\RN{1} in the range 890.2 to 891.7 GHz (ALMA Band 10). The red line represents the synthetic spectrum of CH$_3$NH$_2$ assuming a column density of 2$\times$10$^{17}$ cm$^{-2}$, an excitation temperature of 100 K and a FWHM line width of 3.2 km s$^{-1}$, in a 0$\overset{\arcsec}{.}$26$\times$0$\overset{\arcsec}{.}$26 beam (equivalent to the angular resolution of the data). The abscissa is the rest frequency with respect to the radial velocity towards the region (-7 km s$^{-1}$). 
	The data are shown in black.}
	\label{fig:B10}
\end{figure}

The Band 10 spectrum was acquired as part of project ADS/JAO.ALMA\#2017.1.00717.S. Because the primary beam at Band 10 is only $\sim$7\arcsec, two pointing positions were needed to cover the entire source. Only one of those has been observed - the phase centre was $\alpha$(J2000)~=~17$^{\rm{h}}$20$^{\rm{m}}$53.3$^{\rm{s}}$ $\delta$(J2000)~=~$-35^{\circ}$46\arcmin59.0\arcsec. The spectrum presented in Fig. \ref{fig:B10} was extracted from a position with coordinates (J2000~17$^{\rm{h}}$20$^{\rm{m}}$53.3$^{\rm{s}}$, $-35^{\circ}$46\arcmin59.0\arcsec), chosen off the bright continuum peak of MM1, to minimise the number of transitions driven into absorption. A detailed first look at the data is presented in \citet{McGuire2018}. We present the spectrum here to support the identification of CH$_3$NH$_2$ in NGC 6334\RN{1}, but caution that the excitation conditions and column density in these data at this position are not directly comparable to the Band 7 data discussed in this work. Table \ref{tab:B10_line_summary} lists the catalogue frequencies and other spectroscopic data for the CH$_3$NH$_2$ transitions shown in Fig. \ref{fig:B10}.

\longtab{
	\begin{longtable}{llcccc}
		\caption{\label{tab:B10_line_summary} Summary of the brightest CH$_3$NH$_2$ lines between 890.2 and 891.7 GHz} \\
		\toprule
		\multicolumn{2}{c}{Transition} & Catalogue Frequency & $E_{\textrm{up}}$ & $A_{\textrm{ij}}$ & Catalogue \\
		$[\textrm{QN}]_{\textrm{up}}$
		& $[\textrm{QN}]_{\textrm{low}}$
		& [MHz] & [K] & $\times$10$^{-4}$ [s$^{-1}$] & \\
		\midrule
		\endfirsthead
		\multicolumn{6}{c}{\textit{Continued from previous page}} \\
		\hline
		\multicolumn{2}{c}{Transition} & Catalogue Frequency & $E_{\textrm{up}}$ & $A_{\textrm{ij}}$ & Catalogue \\
		$[\textrm{QN}]_{\textrm{up}}$
		& $[\textrm{QN}]_{\textrm{low}}$
		& [MHz] & [K] & $\times$10$^{-4}$ [s$^{-1}$] & \\
		\midrule
		\endhead
		\midrule
		\multicolumn{6}{c}{\textit{Continued on next page}}\\
		\endfoot
		\bottomrule
		\endlastfoot
		13 6 E1-1 13 & 13 5 E1-1 13 & 890 291.6353 & 333.72 & 8.77 & \cite{Motiyenko2014} \\
		13 6 E1-1 14 & 13 5 E1-1 14 & 890 291.7865 & 333.72 & 9.61 & \\
		13 6 E1-1 12 & 13 5 E1-1 12 & 890 291.7981 & 333.72 & 8.18 & \\
		12 6 E1-1 12 & 12 5 E1-1 12 & 890 383.6647 & 306.08 & 8.57 & \\
		12 6 E1-1 13 & 12 5 E1-1 13 & 890 383.8427 & 306.08 & 9.39 & \\
		12 6 E1-1 11 & 12 5 E1-1 11 & 890 383.8576 & 306.08 & 8.00 & \\
		20 6 B2 20 & 20 5 B1 21 & 890 402.2366 & 589.75 & 9.80 & \\
		20 6 B2 21 & 20 5 B1 21 & 890 402.2883 & 586.75 & 10.3 & \\
		20 6 B2 19 & 20 5 B1 19 & 890 402.2909 & 586.75 & 9.33 & \\
		20 6 B1 20 & 20 5 B2 21 & 890 404.2829 & 589.75 & 9.80 & \\
		20 6 B1 21 & 20 5 B2 21 & 890 404.3348 & 586.75 & 10.3 & \\
		20 6 B1 19 & 20 5 B2 19 & 890 404.3374 & 586.75 & 9.33 & \\	
		11 6 E1-1 11 & 11 5 E1-1 11 & 890 464.2275 & 280.56 & 8.18 & \\
		11 6 E1-1 12 & 11 5 E1-1 12 & 890 464.4391 & 280.56 & 8.97 & \\
		11 6 E1-1 10 & 11 5 E1-1 10 & 890 464.4584 & 280.56 & 7.64 & \\
		10 6 E1-1 10 & 10 5 E1-1 10 & 890 534.3699 & 257.17 & 7.64 & \\
		10 6 E1-1 11 & 10 5 E1-1 11 & 890 534.6247 & 257.17 & 8.57 & \\
		10 6 E1-1 9 & 10 5 E1-1 9 & 890 534.6502 & 257.17 & 6.96 & \\
		9 6 E1-1 9 & 9 5 E1-1 9 & 890 595.0509 & 235.9 & 6.96 & \\
		9 6 E1-1 10 & 9 5 E1-1 10 & 890 595.3625 & 235.9 & 7.81 & \\
		9 6 E1-1 8 & 9 5 E1-1 8 & 890 595.3973 & 235.9 & 6.35 & \\
		8 6 E1-1 8 & 8 5 E1-1 8 & 890 647.1409 & 216.76 & 6.07 & \\
		8 6 E1-1 9 & 8 5 E1-1 9 & 890 647.5298 & 216.76 & 6.96 & \\
		8 6 E1-1 7 & 8 5 E1-1 7 & 890 647.5788 & 216.76 & 5.41 & \\
		7 6 E1-1 7 & 7 5 E1-1 7 & 890 691.4204 & 199.74 & 4.82 & \\
		7 6 E1-1 8 & 7 5 E1-1 8 & 890 691.9184 & 199.74 & 5.53 & \\
		7 6 E1-1 6 & 7 5 E1-1 6 & 890 691.9903 & 199.74 & 4.20 & \\
		6 6 E1-1 6 & 6 5 E1-1 6 & 890 728.5759 & 184.85 & 2.90 & \\
		6 6 E1-1 7 & 6 5 E1-1 7 & 890 729.5759 & 184.85 & 3.49 & \\
		6 6 E1-1 5 & 6 5 E1-1 5 & 890 729.3469 & 184.85 & 2.53 & \\
		18 6 B2 18 & 18 5 B1 18 & 890 757.4027 & 503.88 & 9.61 & \\
		18 6 B2 19 & 18 5 B1 19 & 890 757.4734 & 503.88 & 10.3 & \\
		18 6 B1 18 & 18 5 B2 18 & 890 758.1200 & 503.88 & 9.61 & \\	
		18 6 B1 19 & 18 5 B2 19 & 890 758.1908 & 503.88 & 10.3 & \\		
		11 3 A1 11 & 10 2 A2 10 & 890 875.8080 & 175.87 & 7.94 & \\
		11 3 A1 12 & 10 2 A2 11 & 890 875.9815 & 175.87 & 8.71 & \\
		11 3 A1 10 & 10 2 A2 9 & 890 876.0048 & 175.87 & 7.24 & \\
		17 6 B1 17 & 17 5 B2 17 & 890 904.3890 & 465.64 & 9.61 & \\
		17 6 B1 18 & 17 5 B2 18 & 890 904.4711 & 465.64 & 10.3 & \\
		17 6 B1 16 & 17 5 B2 16 & 890 904.4759 & 465.64 & 8.97 & \\
		17 6 B2 17 & 17 5 B1 17 & 890 904.7947 & 465.64 & 9.61 & \\
		17 6 B2 18 & 17 5 B1 18 & 890 904.8768 & 465.64 & 10.1 & \\
		17 6 B2 16 & 17 5 B1 16 & 890 904.8816 & 465.64 & 8.97 & \\
		16 6 B2 16 & 16 5 B1 16 & 891 033.4246 & 429.50 & 9.39 & \\
		16 6 B2 17 & 16 5 B1 17 & 891 033.5198 & 429.50 & 10.1 & \\
		16 6 B2 15 & 16 5 B1 15 & 891 033.5258 & 429.50 & 8.97 & \\
		16 6 B1 16 & 16 5 B2 16 & 891 033.6460 & 429.50 & 9.39 & \\
		16 6 B1 17 & 16 5 B2 17 & 891 033.7412 & 429.50 & 10.1 & \\
		16 6 B1 15 & 16 5 B2 15 & 891 033.7472 & 429.50 & 8.97 & \\
		15 6 B1 15 & 15 5 B2 15 & 891 146.2027 & 395.48 & 9.18 & \\
		15 6 B1 16 & 15 5 B2 16 & 891 146.3132 & 395.48 & 9.84 & \\
		15 6 B2 15 & 15 5 B1 15 & 891 146.3187 & 395.48 & 9.18 & \\
		15 6 B1 14 & 15 5 B2 14 & 891 146.3206 & 395.48 & 8.77 & \\
		15 6 B2 16 & 15 5 B1 16 & 891 146.4292 & 395.48 & 9.84 & \\
		15 6 B2 14 & 15 5 B1 14 & 891 146.4366 & 395.48 & 8.77 & \\
		14 6 B2 14 & 14 5 B1 14 & 891 244.3081 & 363.58 & 9.18 & \\
		14 6 B1 14 & 14 5 B2 14 & 891 244.3661 & 363.58 & 9.18 & \\
		14 6 B2 15 & 14 5 B1 15 & 891 244.4368 & 363.58 & 9.84 & \\
		14 6 B2 13 & 14 5 B1 13 & 891 244.4460 & 363.58 & 8.57 & \\
		14 6 B1 15 & 14 5 B2 15 & 891 244.4948 & 363.58 & 9.84 & \\
		14 6 B1 13 & 14 5 B2 13 & 891 244.5040 & 363.58 & 8.57 & \\
		13 6 B1 13 & 13 5 B2 13 & 891 329.2209 & 333.82 & 8.77 & \\
		13 6 B2 13 & 13 5 B1 13 & 891 329.2483 & 333.82 & 8.77 & \\
		13 6 B1 14 & 13 5 B2 14 & 891 329.3714 & 333.82 & 9.61 & \\
		13 6 B1 12 & 13 5 B2 12 & 891 329.3831 & 333.82 & 8.18 & \\
		13 6 B2 14 & 13 5 B1 14 & 891 329.3989 & 333.82 & 9.61 & \\
		13 6 B2 12 & 13 5 B1 12 & 891 329.4106 & 333.82 & 8.18 & \\	
		12 6 B2 12 & 12 5 B1 12 & 891 402.3183 & 306.18 & 8.57 & \\
		12 6 B1 12 & 12 5 B2 12 & 891 402.3306 & 306.18 & 8.57 & \\
		12 6 B2 13 & 12 5 B1 13 & 891 402.4958 & 306.18 & 9.39 & \\
		12 6 B1 13 & 12 5 B2 13 & 891 402.5081 & 306.18 & 9.39 & \\
		12 6 B2 11 & 12 5 B1 11 & 891 402.5107 & 306.18 & 8.00 & \\
		12 6 B1 11 & 12 5 B2 11 & 891 402.5229 & 306.18 & 8.00 & \\
		11 6 B1 11 & 11 5 B2 11 & 891 464.8770 & 280.66 & 8.18 & \\
		11 6 B2 11 & 11 5 B1 11 & 891 464.8820 & 280.66 & 8.18 & \\
		11 6 B1 12 & 11 5 B2 12 & 891 465.0882 & 280.66 & 8.97 & \\
		11 6 B2 12 & 11 5 B1 12 & 891 465.0932 & 280.66 & 8.97 & \\
		11 6 B1 10 & 11 5 B2 10 & 891 465.1075 & 280.66 & 7.46 & \\
		11 6 B2 10 & 11 5 B1 10 & 891 465.1125 & 280.66 & 7.46 & \\
		10 6 B2 10 & 10 5 B1 10 & 891 518.0731 & 257.26 & 7.64 & \\
		10 6 B1 10 & 10 5 B2 10 & 891 518.0750 & 257.26 & 7.64 & \\
		10 6 B2 11 & 10 5 B1 11 & 891 518.3275 & 257.26 & 8.57 & \\
		10 6 B1 11 & 10 5 B2 11 & 891 518.3294 & 257.26 & 8.57 & \\
		10 6 B2 9 & 10 5 B1 9 & 891 518.3531 & 257.26 & 6.96 & \\
		10 6 B1 9 & 10 5 B2 9 & 891 518.3550 & 257.26 & 6.96 & \\
		9 6 B1 9 & 9 5 B2 9 & 891 562.9831 & 235.99 & 6.96 & \\
		9 6 B2 9 & 9 5 B1 9 & 891 562.9838 & 235.99 & 6.96 & \\
		9 6 B1 10 & 9 5 B2 10 & 891 563.2945 & 235.99 & 7.81 & \\
		9 6 B2 10 & 9 5 B1 10 & 891 563.2952 & 235.99 & 7.81 & \\
		9 6 B1 8 & 9 5 B2 8 & 891 563.3294 & 235.99 & 6.35 & \\
		9 6 B2 8 & 9 5 B1 8 & 891 563.3300 & 235.99 & 6.35 & \\
		8 6 B2 8 & 8 5 B1 8 & 891 600.5836 & 216.85 & 6.07 & \\
		8 6 B1 8 & 8 5 B2 8 & 891 600.5838 & 216.85 & 6.07 & \\
		8 6 B2 9 & 8 5 B1 9 & 891 600.9724 & 216.85 & 6.96 & \\
		8 6 B1 9 & 8 5 B2 9 & 891 600.9725 & 216.85 & 6.96 & \\
		8 6 B2 7 & 8 5 B1 7 & 891 601.0213 & 216.85 & 5.41 & \\
		8 6 B1 7 & 8 5 B2 7 & 891 601.0215 & 216.85 & 5.41 & \\
		7 6 B1 7 & 7 5 B2 7 & 891 631.7493 & 199.83 & 4.82 & \\
		7 6 B2 7 & 7 5 B1 7 & 891 631.7493 & 199.83 & 4.82 & \\	
		7 6 B1 8 & 7 5 B2 8 & 891 632.2472 & 199.83 & 5.66 & \\	
		7 6 B2 8 & 7 5 B1 8 & 891 632.2473 & 199.83 & 5.66 & \\	
		7 6 B1 6 & 7 5 B2 6 & 891 632.3190 & 199.83 & 4.29 & \\	
		7 6 B2 6 & 7 5 B1 6 & 891 632.3191 & 199.83 & 4.29 & \\	
		6 6 B1 6 & 6 5 B2 6 & 891 657.2500 & 184.94 & 2.97 & \\
		6 6 B2 6 & 6 5 B1 6 & 891 657.2500 & 184.94 & 2.97 & \\	
		6 6 B1 7 & 6 5 B2 7 & 891 657.9096 & 184.94 & 3.49 & \\
		6 6 B2 7 & 6 5 B1 7 & 891 657.9096 & 184.94 & 3.49 & \\
		6 6 B1 5 & 6 5 B2 5 & 891 658.0209 & 184.94 & 2.53 & \\
		6 6 B2 5 & 6 5 B1 5 & 891 658.0209 & 184.94 & 2.53 & \\	
		\bottomrule
	\end{longtable}
	\tablefoot{Quantum numbers are J $\textrm{K}_{\textrm{a}}$ $\Gamma$ F, following the notation of \citet{Motiyenko2014}.}
}
\clearpage
\twocolumn

\section{Methanimine, methyl cyanide and formamide} \label{app:results}
In this appendix, the detections of CH$_2$NH, the $^{13}$C- and $^{15}$N-methyl cyanide isotopologues and NH$_2$CHO are discussed in detail. The detected lines and best-fit models are shown in Figs. \ref{fig:13CH3CNlines}--\ref{fig:NH2CHOlines}.

\subsection{Methanimine CH$_2$NH}
A single (hyperfine split) transition of CH$_2$NH is covered by the data. The CH$_2$NH transition, located at 302.565 GHz, and best-fit synthetic spectrum for each of the regions are included in Fig. \ref{fig:NH2CHOlines}. Unfortunately, the CH$_2$NH feature is situated in the wing of a much stronger transition, located at approximately 302.562~GHz, identified as CH$_3$OCHO. It should be noted however, that the peak in the data at 302.562 GHz is only partly reproduced by the synthetic spectrum of CH$_3$OCHO and additional contributions to the peak from other species, which are not included in the JPL or CDMS molecular databases, can therefore not be excluded. Because of this blend, we report only upper limits on the column density of CH$_2$NH in each of the studied regions. The CH$_2$NH column densities are $\leq$5.2$\times$10$^{16}$ cm$^{-2}$ assuming $T_{\textrm{ex}}$= 215K for MM1~\RN{2}, $\leq$5.0$\times$10$^{16}$ cm$^{-2}$ assuming $T_{\textrm{ex}}$ = 165 K for MM2~\RN{1} and $\leq$10$^{15}$ cm$^{-2}$ assuming $T_{\textrm{ex}}$~=~120 K for MM3~\RN{1}. 

\subsection{Methyl cyanide CH$_3$CN} \label{subsec:ch3cn}
\begin{figure*}
	\centering
	\includegraphics[width=1.\textwidth, trim={0.5cm 0 0.5cm 0}, clip]{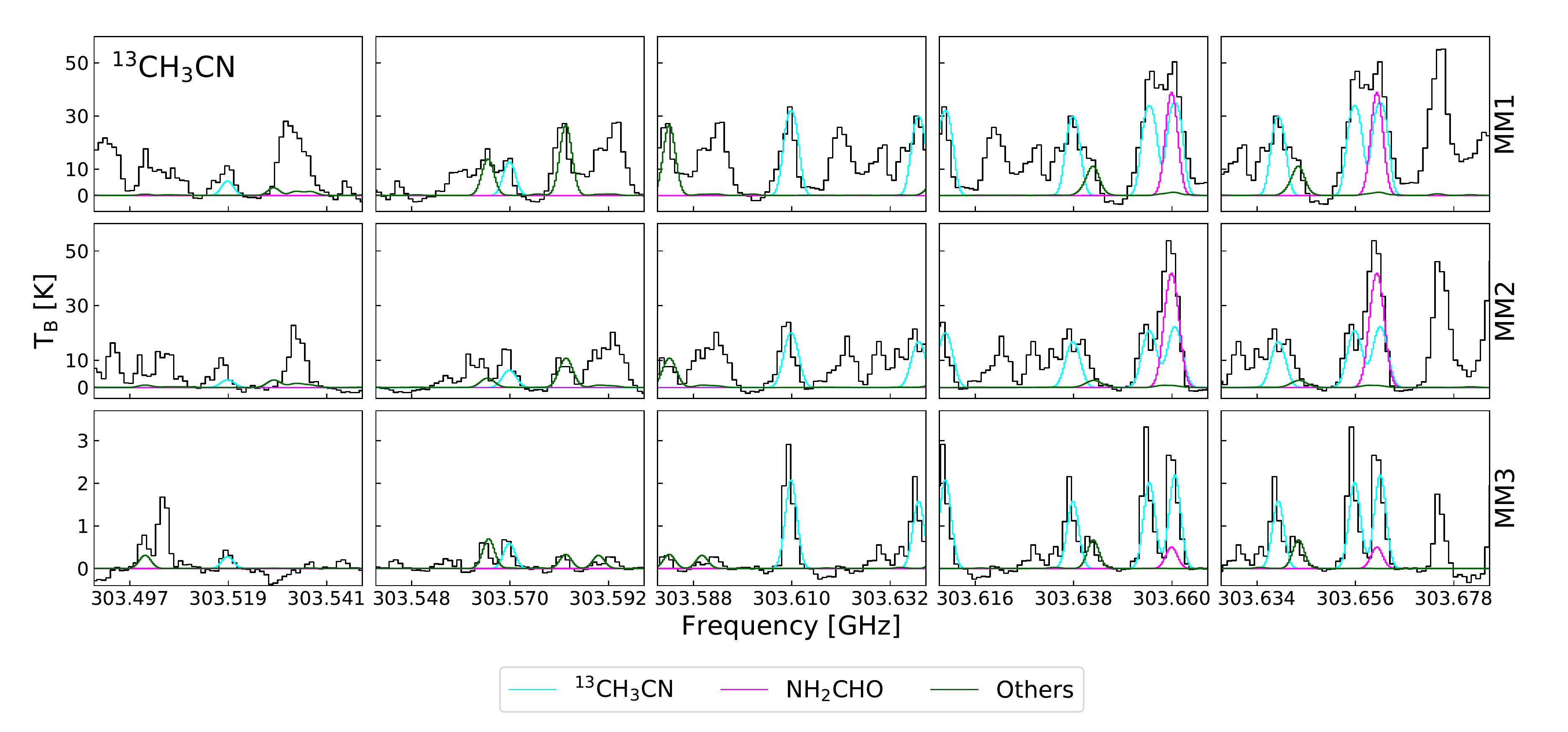} 
	\caption[]{$^{13}$CH$_3$CN transitions detected towards NGC6334\RN{1}. Turquoise, magenta and green lines represent the synthetic spectrum of $^{13}$CH$_3$CN and NH$_2$CHO and the sum of spectra of other contributing species respectively. The abscissa is the rest frequency with respect to the radial velocity towards each of the hot cores (listed in Table \ref{tab:models}). 
	The data are shown in black. \textit{Top panels}: MM1~\RN{2}. \textit{Middle panels}: MM2~\RN{1}. \textit{Bottom panels}: MM3~\RN{1}.}
	\label{fig:13CH3CNlines}
\end{figure*}

\begin{figure*}
	\centering
	\includegraphics[width=1.\textwidth, trim={0.5cm 0 0.5cm 0}, clip]{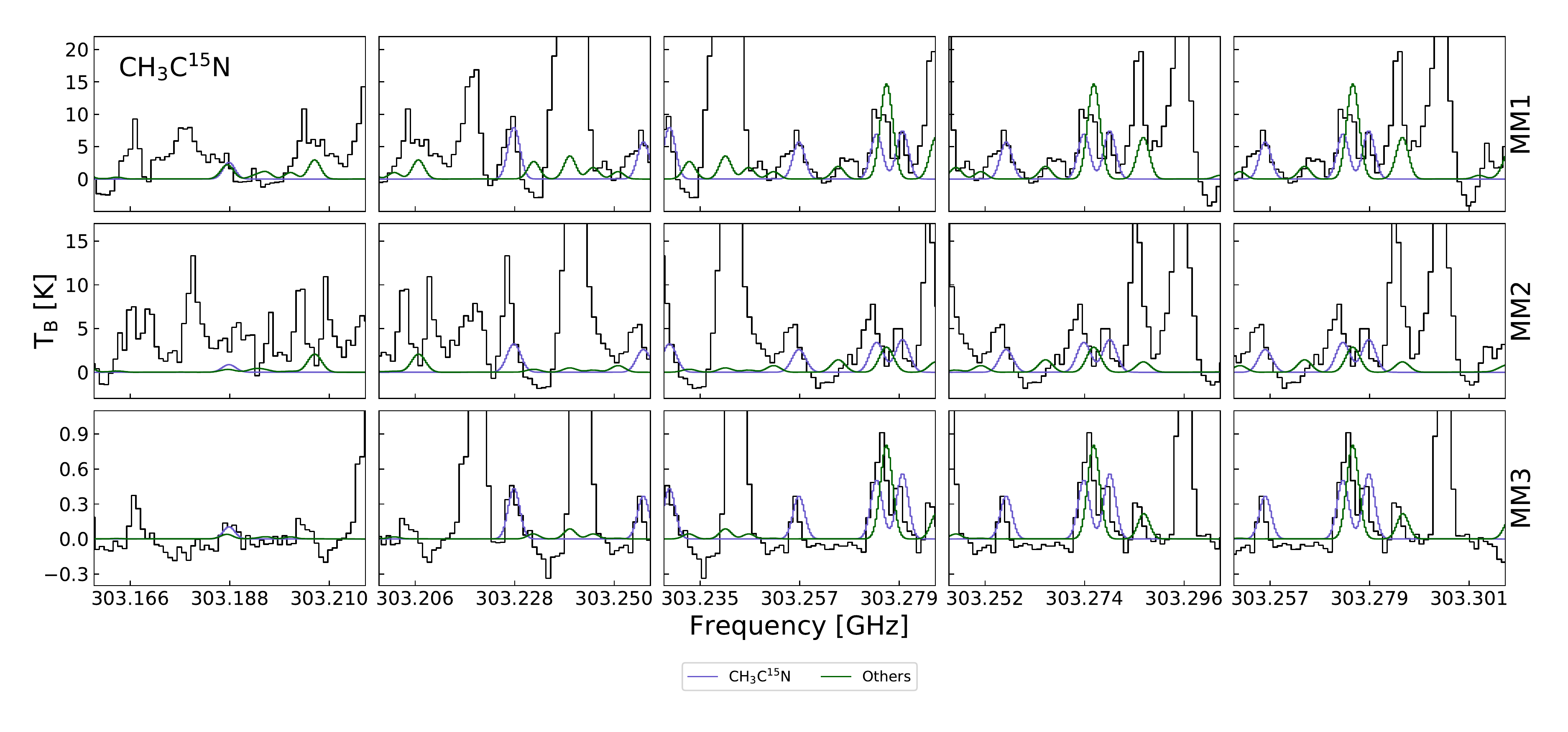} 
	\caption[]{CH$_3$C$^{15}$N transitions detected towards NGC6334\RN{1}. Blue and green lines represent the synthetic spectrum of CH$_3$C$^{15}$N and the sum of spectra of other contributing species respectively. The abscissa is the rest frequency with respect to the radial velocity towards each of the hot cores (listed in Table \ref{tab:models}). 
	The data are shown in black. \textit{Top panels}: MM1~\RN{2}. \textit{Middle panels}: MM2~\RN{1}. \textit{Bottom panels}: MM3~\RN{1}.}
	\label{fig:CH3C15Nlines}
\end{figure*}

There are no transitions of the main CH$_3$CN isotopologue covered by the observations but six transitions belonging to the $^{13}$C- and five transitions belonging to the $^{15}$N-methyl cyanide isotopologues are within the data range. Based on these, the column density of the main CH$_3$CN isotopologue is derived assuming a $^{12}$C/$^{13}$C value of 62 and a $^{14}$N/$^{15}$N value of 422, both derived assuming $d_{\textrm{GC}}$ = 7.02 kpc and the $^{12}$C/$^{13}$C and $^{14}$N/$^{15}$N relations presented by \cite{Milam2005} and \cite{Wilson1999}, respectively. The detected transitions of both $^{13}$CH$_3$CN and CH$_3$C$^{15}$N belong to the J = 17$\rightarrow$16 series around 303.6 GHz and 303.2 GHz, respectively, and have upper state energies in the range 131 to 310 K. No transitions of the $^{13}$C-methyl cyanide isomer CH$_3^{13}$CN are covered by the data. Figures \ref{fig:13CH3CNlines} and \ref{fig:CH3C15Nlines} show the detected methyl cyanide transitions and best-fit models.

\textbf{MM1~\RN{2}:} In the case of MM1~\RN{2}, the best-fit methyl cyanide column densities and excitation temperature are 3.4$\times10^{15}$ cm$^{-2}$ and 70 K for $^{13}$CH$_3$CN, respectively, and 3.3$\times10^{14}$ cm$^{-2}$ and 110 K for CH$_3$C$^{15}$N, respectively. For $^{13}$CH$_3$CN, the uncertainty on $N_{\textrm{s}}$ and $T_{\textrm{ex}}$ is approximately 30\% and 15\%, receptively, while for CH$_3$C$^{15}$N, the approximate uncertainty is 15\% and 45\%, respectively. The ratio of the column densities of $^{13}$CH$_3$CN to CH$_3$C$^{15}$N is a factor of 10, higher than the expected value of 6.8 based on the $^{12}$C/$^{13}$C and $^{14}$N/$^{15}$N relations. Of the six detected transitions belonging to $^{13}$CH$_3$CN, five are largely uncontaminated by emission from other species and can be assigned distinct peaks in the data. The final transition, located at 303.661 GHz, is blended with a transition of NH$_2$CHO (to be discussed in detail below). The sum of the best-fit spectra of $^{13}$CH$_3$CN and NH$_2$CHO results in a model with a peak that is $\sim$45\% more intense than the data at 303.661 GHz. Optimising the column density of the $^{13}$CH$_3$CN model to fit the blended transition results in underestimated model peak intensities for the remaining $^{13}$CH$_3$CN lines with respect to the data. The best-fit column density optimised to the blended line is a factor two lower than the best-fit column density when optimising to all $^{13}$CH$_3$CN transitions. In contrast to $^{13}$CH$_3$CN, the transitions of CH$_3$C$^{15}$N are mostly blended and only two of the five transitions, located at 303.228 GHz and 303.257 GHz, can be assigned distinct counterparts in the data. The remaining transitions are blended with transitions of CH$_3$SH located around 303.187 GHz and 303.276 GHz respectively. Excluding the contribution from CH$_3$SH to the model does not change the value of the best-fit column density of CH$_3$C$^{15}$N but it should be noted that the summed best-fit spectra of CH$_3$C$^{15}$N and CH$_3$SH overshoot the data feature located at 303.276 GHz with approximately 50\%. 

\textbf{MM2~\RN{1}}: The best-fit excitation temperature and column density for $^{13}$CH$_3$CN in the MM2~\RN{1} region is 80~K and 1.4$\times10^{15}$ cm$^{-2}$, respectively. For CH$_3$C$^{15}$N, the excitation is not well contained and therefore the best-fit excitation temperature for $^{13}$CH$_3$CN is adopted. For this temperature, the best-fit column density is 1.8$\times10^{14}$ cm$^{-2}$. The uncertainty on $N_{\textrm{s}}$ and $T_{\textrm{ex}}$ is approximately 30\% in the case of $^{13}$CH$_3$CN and 20\% in the case of CH$_3$C$^{15}$N. As is the case of MM1~\RN{2}, a single $^{13}$CH$_3$CN line and three CH$_3$C$^{15}$N lines are contaminated by emission from NH$_2$CHO and CH$_3$SH respectively. However, while the best-fit $^{13}$CH$_3$CN column density optimised purely based on the unblended transitions is higher by $\sim$20\% with respect to the value derived when all transitions are included, the best-fit column density of CH$_3$C$^{15}$N remains the same. The sum of spectra of $^{13}$CH$_3$CN and NH$_2$CHO results in a model peak which is 15\% brighter than the data at the location of the blended $^{13}$CH$_3$CN line while including the contribution from CH$_3$SH to the model of CH$_3$C$^{15}$N has only little effect on the summed spectra. The $^{13}$CH$_3$CN to CH$_3$C$^{15}$N ratio is lower than the ratio derived for MM1~\RN{2} and has a value of 7.8.

\textbf{MM3~\RN{1}:} The best-fit methyl cyanide column densities in region MM3~\RN{1} are lower than for both MM1~\RN{2} and MM2~\RN{1} with values of 9$\times10^{13}$ cm$^{-2}$ for $^{13}$CH$_3$CN and 2.3$\times10^{13}$ cm$^{-2}$ for CH$_3$C$^{15}$N, with excitation temperatures of 90 and 70~K, respectively. The uncertainty on $N_{\textrm{s}}$ and $T_{\textrm{ex}}$ is approximately 10\% and 20\%, respectively, for $^{13}$CH$_3$CN and 30\% and a factor of two, respectively, for CH$_3$C$^{15}$N. The ratio of the $^{13}$CH$_3$CN to CH$_3$C$^{15}$N column density is a factor of 3.9, lower than the expected value. For $^{13}$CH$_3$CN as well as for CH$_3$C$^{15}$N, the best-fit column density remains unchanged when contributions from blending species are included.   

\subsection{Formamide NH$_2$CHO}
\begin{figure*}[h]
	\centering
	\includegraphics[width=1.\textwidth, trim={0.5cm 0 0.8cm 0}, clip]{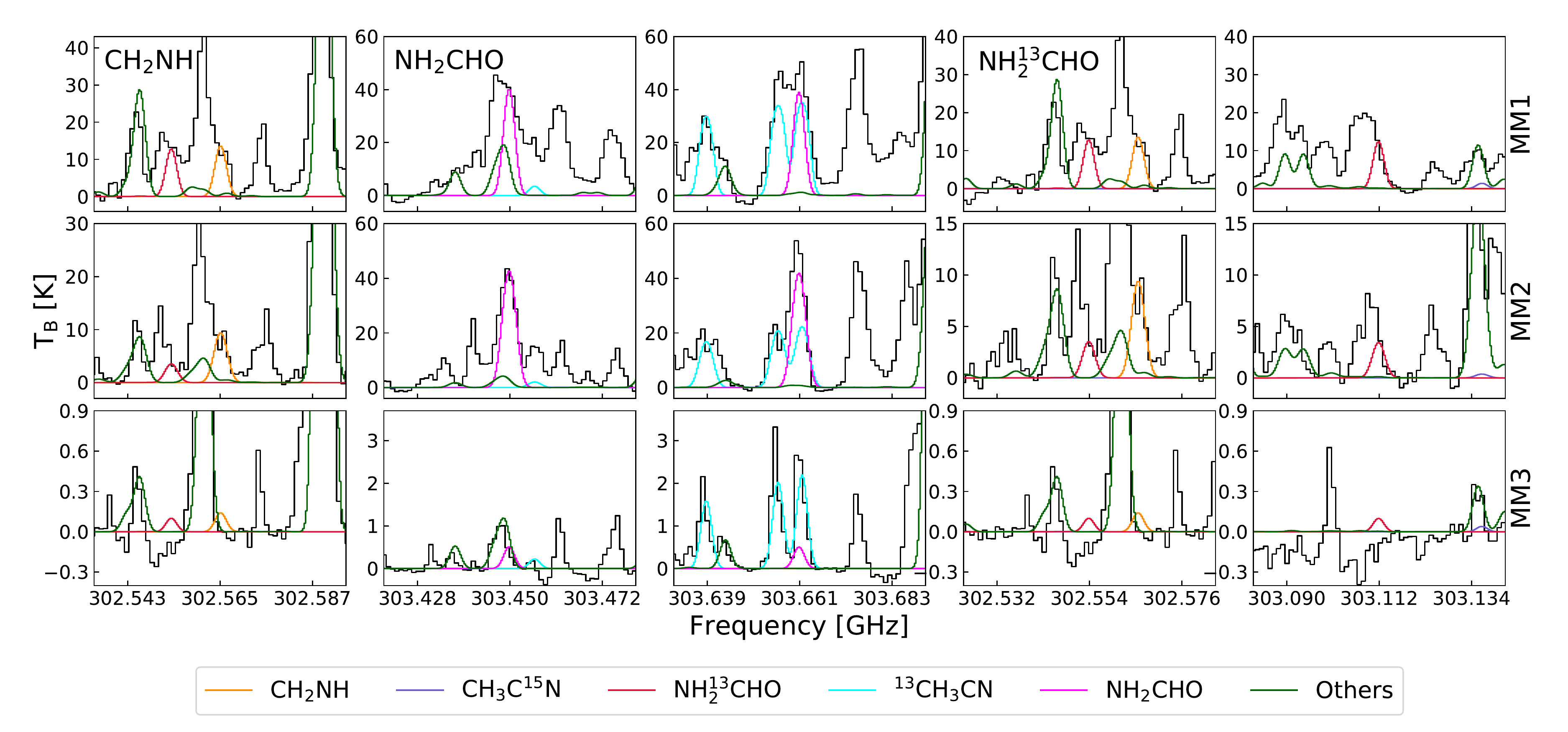} 
	\caption[]{CH$_2$NH, NH$_2$CHO and NH$_2^{13}$CHO transitions detected towards NGC6334\RN{1}. Orange, magenta, red and green lines represent the synthetic spectrum of CH$_2$NH, NH$_2$CHO and NH$_2^{13}$CHO and the sum of spectra of other contributing species respectively. The abscissa is the rest frequency with respect to the radial velocity towards each of the hot cores (listed in Table \ref{tab:models}). The data are shown in black. \textit{Top panels}: MM1~\RN{2}. \textit{Middle panels}: MM2~\RN{1}. \textit{Bottom panels}: MM3~\RN{1}.}
	\label{fig:NH2CHOlines}
\end{figure*}

While a total of 18 NH$_2$CHO transitions are covered by the data, only two are bright enough to be detected towards the NGC~6334\RN{1} region. As discussed above, one of these lines, located at 303.661 GHz, is blended with a transition of $^{13}$CH$_3$CN. The second transition, located at 303.450 GHz, is also blended but with emission from CH$_3$SH. Fig. \ref{fig:NH2CHOlines} shows the transitions and the best-fit model for each of the regions. In addition to the main NH$_2$CHO isotopologue, two transitions of NH$_2^{13}$CHO are within the data range. Transitions belonging to the $^{15}$N- and deuterated formamide isotopologues are too weak to be detected.

\textbf{MM1~\RN{2}:} The synthetic spectrum that best reproduces the NH$_2$CHO lines detected towards the MM1~\RN{2} region has a column density of 7.0$\times10^{15}$ cm$^{-2}$, assuming an excitation temperature of 215 K. The uncertainty on $N_{\textrm{s}}$ is approximately 25\%. In Sect. \ref{subsec:ch3cn} the NH$_2$CHO line blended with $^{13}$CH$_3$CN was discussed and it was concluded that the sum of the optimised $^{13}$CH$_3$CN and NH$_2$CHO spectra results in a modelled spectrum which overshoots the data by approximately 45\%. The second detected NH$_2$CHO transition is blended with CH$_3$SH. However, the contribution from this species to the data feature at the location of the NH$_2$CHO transition is small. Excluding the contribution from CH$_3$SH to the full spectrum, results in a best-fit NH$_2$CHO column density which is less than a factor of two higher than the best-fit value which includes the blending species. For NH$_2^{13}$CHO the best-fit column density is $\leq$2.0$\times10^{15}$ cm$^{-2}$ assuming an excitation temperature of 215 K. Both the detected transitions of NH$_2^{13}$CHO are located in the wing of brighter emission lines and therefore, as in the case of CH$_2$NH, the best-fit model parameters are listed as upper limits only. Neither of the features blended with the NH$_2^{13}$CHO transitions are reproduced by the modelled spectra of the potential blending species listed in Table \ref{tab:main_blending_species}. 

\textbf{MM2~\RN{1}:} For region MM2~\RN{1} the best-fit column densities are 7.6$\times10^{15}$ cm$^{-2}$ and $\leq$5.0$\times10^{14}$ cm$^{-2}$ for NH$_2$CHO and NH$_2^{13}$CHO respectively, both assuming an excitation temperature of 165 K. The uncertainty on the column density of NH$_2$CHO is $\sim$10\%. As in the case of MM1~\RN{2}, both the blended NH$_2$CHO transitions are slightly overproduced with respect to the data when the modelled spectra of $^{13}$CH$_3$CN and other blending species are included in the fit. Excluding these blending species however, only increases the best-fit NH$_2$CHO column density by 10\%. As in the case of MM1~\RN{2}, the NH$_2^{13}$CHO column density is listed as an upper limit.

\textbf{MM3~\RN{1}:} In region MM3~\RN{1} only the transitions of the main NH$_2$CHO isotopologue are detected. However, since the data features at the locations of the NH$_2$CHO transitions can also be reproduced by the respective blending species $^{13}$CH$_3$CN and CH$_3$SH, the detection of NH$_2$CHO in this region is tentative and its column density reported as an upper limit. The best-fit modelled spectrum of NH$_2$CHO has a column density of $\leq$5.0$\times10^{13}$ cm$^{-2}$ assumes an excitation temperature of 120 K.

\onecolumn
\section{Potential blending species}
In this appendix, a list of the potential blending species and the column densities and excitation temperatures used to fit them, is present. The species have transitions which overlap in frequency with transitions of CH$_3$NH$_2$, CH$_2$NH, NH$_2$CHO or the CH$_3$CN isotopologues and may therefore be contributing to the observed spectrum extracted from each of the studied regions.
 
\begin{table}[h]
	\centering
	\caption{Model parameters of potential blending species}
	\label{tab:main_blending_species}
	\begin{tabular}{!{\extracolsep{4pt}}lccccccc!{}}
		\toprule
		Species & Catalogue & \multicolumn{2}{c}{MM1~\RN{2}} & \multicolumn{2}{c}{MM2~\RN{1}} & \multicolumn{2}{c}{MM3~\RN{1}} \\
		\cline{3-4}
		\cline{5-6}
		\cline{7-8}
		& & $N_{\textrm{s}}$ [cm$^{-2}$] & $T_{\textrm{ex}}$ [K] & $N_{\textrm{s}}$ [cm$^{-2}$] & $T_{\textrm{ex}}$ [K] & $N_{\textrm{s}}$ [cm$^{-2}$] & $T_{\textrm{ex}}$ [K] \\
		\midrule
		$^{13}$CH$_3$OH\tablefootmark{a} & CDMS & 7.4$\times 10^{17}$ & [215] & 6.6$\times 10^{17}$ & [165] & 9.0$\times 10^{16}$ & [120] \\
		CH$_3^{18}$OH\tablefootmark{a} & CDMS & 2.0$\times 10^{17}$ & [215] & 8.0$\times 10^{16}$ & [165] & 1.4$\times 10^{16}$ & [120] \\
		C$_2$H$_5$OH & JPL & 5.0$\times 10^{17}$ & [215] & 2.0$\times 10^{17}$ & [300] & 6.0$\times 10^{15}$ & [200] \\
		CH$_3$CH$_2$OD & CDMS & 5.0$\times 10^{15}$ & [215] & 6.0$\times 10^{15}$ & [165] & 6.0$\times 10^{14}$ & [50] \\
		CH$_2$DCH$_2$OH & CDMS & 5.0$\times 10^{15}$ & [215] & 3.0$\times 10^{15}$ & [165] & 2.0$\times 10^{14}$ & [50] \\
		CH$_3$SH & CDMS & 3.0$\times 10^{16}$ & [215] & $<$1.0$\times 10^{16}$ & [165] & 1.0$\times 10^{15}$ & [120] \\
		C$_2$H$_3$CN & CDMS & 1.0$\times 10^{15}$ & [215] & $<$1.0$\times 10^{15}$ & [165] & -- & -- \\
		$^{13}$CH$_2$CHCN & CDMS & 8.0$\times 10^{14}$ & [215] & 2.0$\times 10^{14}$ & [165] & -- & -- \\
		CH$_3$COCH$_3$ & JPL & 1.5$\times 10^{17}$ & [215] & 2.0$\times 10^{17}$ & [165] & 2.0$\times 10^{15}$ & [120] \\
		CH$_3$OCHO & JPL & 1.2$\times 10^{17}$ & [215] & 2.5$\times 10^{17}$ & [165] & 9.0$\times 10^{16}$ & [120] \\
		HCOCH$_2$OH & JPL & 3.5$\times 10^{16}$ & [215] & 8.0$\times 10^{15}$ & [165] & -- & -- \\
		C$_2$H$_5$SH & CDMS & 3.0$\times 10^{16}$ & [215] & 4.0$\times 10^{16}$ & [165] & 7.0$\times 10^{14}$ & [120] \\
		CH$_3$C$_3$N & CDMS & 1.0$\times 10^{15}$ & [215] & 5.0$\times 10^{14}$ & [165] & 5.0$\times 10^{13}$ & [120] \\
		C$_3$H$_7$CN & CDMS & 5.0$\times 10^{15}$ & [215] & 5.0$\times 10^{15}$ & [165] & -- & -- \\
		\bottomrule
	\end{tabular}
	\tablefoot{All models assume $\theta_{\textrm{s}}$ = 1$\overset{\second}{.}$0, $v_{\textrm{LSR}}$ = -6.7 km s$^{-1}$ for MM1~\RN{2} and $v_{\textrm{LSR}}$ = -9.0 km s$^{-1}$ for MM2~\RN{1} and MM3~\RN{1}, and FWHM = 3.5 km s$^{-1}$ for MM2~\RN{1} and FWHM = 3.0 km s$^{-1}$ for MM1~\RN{2} and MM3~\RN{1}. $N_{\textrm{s}}$ is the highest value consistent with the data. \tablefoottext{a}{Values from \cite{Bogelund2018}}}
\end{table}

\end{document}